\documentclass[11pt]{article}

\usepackage[T1]{fontenc}
\usepackage[utf8]{inputenc}
\usepackage{lmodern}
\usepackage[margin=1in]{geometry}
\usepackage{amsmath,amssymb,amsthm,mathtools}
\usepackage{booktabs,tabularx,array}
\usepackage{microtype}
\usepackage[hidelinks]{hyperref}
\hypersetup{pdftitle={Bounded Metrics for Probability Measures on the Sphere},pdfsubject={Tutorial review of bounded metrics for spherical probability measures},pdfkeywords={Wasserstein, spherical measures, MMD, energy distance, Hellinger, Jensen-Shannon, Hilbertian metric}}
\usepackage[round,authoryear]{natbib}

\newcommand{\doi}[1]{\href{https://doi.org/#1}{doi:\nolinkurl{#1}}}
\usepackage{enumitem}
\usepackage{placeins}
\usepackage{float}
\usepackage{graphicx}
\usepackage[font=small,labelfont=bf]{caption}

\usepackage{tikz}
\usetikzlibrary{arrows.meta,calc,positioning,backgrounds}
\numberwithin{equation}{section}
\newtheorem{theorem}{Theorem}[section]
\newtheorem{proposition}[theorem]{Proposition}
\newtheorem{lemma}[theorem]{Lemma}

\theoremstyle{definition}

\newtheorem{example}[theorem]{Example}
\theoremstyle{remark}

\newcommand{\R}{\mathbb R}
\newcommand{\Sph}{\mathbb S^{d-1}}
\newcommand{\Prob}{\mathcal P(\Sph)}
\newcommand{\E}{\mathbb E}
\newcommand{\dd}{\,\mathrm d}
\newcommand{\1}{\mathbf 1}

\newcommand{\SW}{\operatorname{SW}_2}
\newcommand{\SWL}{\operatorname{SW}_{2,L}}
\newcommand{\MMD}{\operatorname{MMD}}
\newcommand{\TV}{\operatorname{TV}}
\newcommand{\JS}{\operatorname{JS}}
\newcommand{\KL}{\operatorname{KL}}
\newcommand{\Cot}{C_{\mathrm{OT}}}
\newcommand{\Aot}{A_{\mathrm{OT}}}
\newcommand{\Csw}{C_{\mathrm{SW}}}
\newcommand{\Dsw}{D_{\mathrm{SW}}}
\newcommand{\Asw}{A_{\mathrm{SW}}}
\newcommand{\DE}{D_{\mathrm E}}
\newcommand{\Energy}{\mathcal E_{\mathrm c}}

\newcommand{\ip}[2]{\langle #1,#2\rangle}
\newcommand{\norm}[1]{\lVert #1\rVert}

\newcolumntype{Y}{>{\raggedright\arraybackslash}X}

\title{Bounded Metrics for Probability Measures on the Sphere}
\author{Kisung You}
\date{}

\begin{document}
\maketitle
\vspace{-1.7em}
\begin{abstract}
Directional distributions can have the same normalized mean direction while differing substantially in shape. Comparing them requires a distance whose response to geometric displacement and probability overlap matches the statistical question. We examine ten bounded metrics or metric families for probability measures on the sphere, scaling each to have maximum value one, covering  transport, sliced transport, kernel and energy distances, and comparisons based on probability mass. We explain which metrics describe weak convergence, which admit Hilbert-space norm representations, and which pairs attain the maximum distance. Examples and counterexamples show how these distinctions affect empirical comparisons and numerical approximations. A summary table and relationship diagram support the choice of a distance. The discussion separates exact metrics from finite representations, unbiased estimators, and regularized transport costs.
\end{abstract}

\section{Introduction}\label{sec:setting}
When two directional distributions have the same normalized mean direction, how should we compare their shapes? Electron scattering provides one motivation where the angular distribution records differences that a normalized average direction can miss \citep{jablonski_2004_ComparisonElectronElasticScattering}. Comparing normalized angular patterns requires a common coordinate system and appropriate treatment of detector sensitivity. The comparison concerns the distribution of directions after normalizing by total scattering intensity.

Comparing full distributions still leaves a choice about what should count as a small difference. Moving probability through a small angle can give a small transport distance, while distances based on exact probability overlap can remain maximal. For equally weighted samples of the same size, nearby one-to-one matches can give small transport distances even when the samples share no locations. A common numerical range alone cannot resolve these differences.

This article reviews ten existing bounded metrics or metric families on the sphere, placing them on the same scale and explaining their relationships. We compare the convergence they describe, their representations as Hilbert norm distances, and the distributions that attain their maximum. Worked examples clarify the consequences for comparing samples, and the discussion of computation distinguishes exact empirical distances from approximations and statistical estimators. The contribution is an organized comparison of established constructions, with no experimental data analysis or claim of a universally preferred distance.

General comparisons of probability metrics appear in \citet{gibbs_2002_ChoosingBoundingProbability}. Background on transport is given by \citet{villani_2009_OptimalTransportOld,santambrogio_2015_OptimalTransportApplied,peyre_2019_ComputationalOptimalTransport}, with a statistical account in \citet{panaretos_2019_StatisticalAspectsWasserstein}. Kernel representations are reviewed by \citet{sriperumbudur_2010_HilbertSpaceEmbeddings,muandet_2017_KernelMeanEmbedding}. Our common spherical setting makes the resulting comparisons explicit for both discrete and continuous distributions.

The rest of this paper unfolds as follows. Section~\ref{sec:table} introduces the setting and summarizes the metrics in a table and diagram. Sections~\ref{sec:wasserstein} to~\ref{sec:js} explain the constructions, followed by their shared relationships and the consequences of numerical approximation. The main text assumes familiarity with linear algebra, calculus, and probability. All proofs are deferred to Appendix.

\section{Setting and comparison of the metrics}\label{sec:table}
\subsection{Measures and ground distances}

For unit vectors $x,y\in\Sph\subset\R^d$, cosine similarity is $x^\top y$. The \emph{chordal distance} measures straight-line separation in the surrounding space. The \emph{geodesic distance} measures the angle between the vectors. They are given by
\begin{equation*}
 \rho_{\mathrm c}(x,y)=\norm{x-y}=\sqrt{2-2x^\top y},
 \qquad
 \rho_{\mathrm g}(x,y)=\arccos(x^\top y).
\end{equation*}
On the unit sphere, the angle in radians is also the length of the shortest path along the surface. Cosine similarity is not a metric because its value is one, rather than zero, when the vectors agree. For distributions, we will likewise distinguish similarities from distances and check the properties of each construction separately.

Write $\Prob$ for all probability distributions on $\Sph$, with $d\geq2$. Formally, these are Borel probability measures, defined on the measurable sets generated by the open subsets of the sphere. This setting includes weighted empirical measures
\begin{equation}\label{eq:empirical}
 \mu=\sum_{i=1}^n a_i\delta_{x_i},\qquad
 a_i\geq0,\quad \sum_i a_i=1,\quad x_i\in\Sph.
\end{equation}
Here $\delta_x$ denotes unit probability mass at $x$, and $a_i$ is the probability assigned to $x_i$. Reordering the points, or splitting one weight among repeated copies of the same point, does not change the measure. All comparisons use the same coordinate system and fixed parameter choices. Treating opposite directions as the same, or rotating each distribution separately before comparison, would be a different problem.

\paragraph{An average can miss shape.}
Let $0<\alpha<\pi/2$, set $z=(0,0,1)$, and put $u_\pm=(\pm\sin\alpha,0,\cos\alpha)$. Compare $\mu=\tfrac12(\delta_{u_+}+\delta_{u_-})$ with $\nu=\delta_z$. Their average vectors are $\cos\alpha\,z$ and $z$, so their normalized mean directions agree, although every metric on probability measures distinguishes the two distributions. The normalization hides the difference, since the unnormalized means differ in length. These point masses illustrate the information lost by normalizing an average, without modeling a particular scattering experiment.

\subsection{Notation}

A \emph{coupling} of $\mu$ and $\nu$ is a joint distribution for a pair $(X,Y)$ such that $X$ has distribution $\mu$ and $Y$ has distribution $\nu$. For finite measures of the form~\eqref{eq:empirical}, it specifies how much mass at each point of one measure is paired with each point of the other. We write $\Pi(\mu,\nu)$ for the set of all couplings. The product measure $\mu\otimes\nu$ is the coupling in which $X$ and $Y$ are independent.

For a measurable map $T$ and $X$ with distribution $\mu$, the distribution of $T(X)$ is the \emph{pushforward} $T_\#\mu$. We write $\mu_n\Rightarrow\mu$ for \emph{weak convergence}. On the compact sphere, this means that $\int h\,d\mu_n\to\int h\,d\mu$ for every continuous real-valued function $h$. In other words, averages of continuous functions converge. The term \emph{topology} will describe which sequences converge in this sense or in a specified distance. A standard reference for weak convergence and compactness of probability measures is \citet{billingsley_1999_ConvergenceProbabilityMeasures}.

The symbol $\sigma$ always denotes the \emph{uniform probability measure} on the sphere, with total mass one. On $\mathbb S^2$, ordinary solid-angle measure $d\Omega$ has total mass $4\pi$, so $d\sigma=d\Omega/(4\pi)$. Thus a density $q$ normalized by $\Sigma=\int q\,d\Omega$ is $q/\Sigma$ relative to $d\Omega$, but $4\pi q/\Sigma$ relative to $d\sigma$. This convention matters for the constants in sliced Wasserstein distance.

Two measures are \emph{mutually singular}, written $\mu\perp\nu$, if they put all their probability on disjoint measurable sets. Equivalently, some measurable set $A$ satisfies $\mu(A)=1$ and $\nu(A)=0$. Distinct point masses provide a simple example. Mutual singularity does not require their closed supports to be disjoint.

When densities are needed, we choose a common finite reference measure $r$. It must \emph{dominate} the measures under discussion, meaning that a set with zero $r$-mass also has zero mass under each of them. Write $f=d\mu/dr$ and $g=d\nu/dr$. These are densities relative to $r$, also called Radon-Nikodym derivatives. They need not be smooth or be densities relative to surface area. For a pair we can always take $r=\mu+\nu$, and for a finite collection we can take its sum. All logarithms are natural.

\subsection{Comparison criteria}

\begin{figure}[p]
\centering
\makebox[\textwidth][c]{%
  \resizebox{\dimexpr\textwidth+0.8in\relax}{!}{%
\definecolor{ink}{HTML}{203040}
\definecolor{muted}{HTML}{5E6874}
\definecolor{bound}{HTML}{345777}
\definecolor{exact}{HTML}{20736B}
\definecolor{weakfill}{HTML}{F3F7FA}
\definecolor{strongfill}{HTML}{FBF7EE}
\definecolor{panelborder}{HTML}{C7D2DC}
\definecolor{strongborder}{HTML}{DACDB5}
\definecolor{hilbert}{HTML}{20736B}
\begin{tikzpicture}[
  x=1cm,y=1cm,
  font=\rmfamily\fontsize{10.5}{13}\selectfont,
  text=ink,
  metric/.style={draw=panelborder,fill=white,rounded corners=3pt,
    line width=.65pt,minimum height=1.44cm,text width=4.42cm,
    inner xsep=6pt,inner ysep=6pt,align=center},
  massmetric/.style={metric,draw=strongborder},
  boundedge/.style={draw=bound,line width=.9pt,-{Latex[length=2.2mm,width=1.5mm]}},
  boundlink/.style={draw=bound,line width=.9pt},
  exactedge/.style={draw=exact,line width=1pt,dash pattern=on 4pt off 2.6pt,
    {Latex[length=2.1mm,width=1.5mm]}-{Latex[length=2.1mm,width=1.5mm]}},
  mixedge/.style={draw=exact,line width=1pt,dash pattern=on 4pt off 2.6pt,
    -{Latex[length=2.1mm,width=1.5mm]}},
  edgeword/.style={font=\rmfamily\fontsize{9.5}{12}\selectfont,align=center,
    inner sep=3pt,text=bound},
  exactword/.style={edgeword,text=exact},
  badge/.style={circle,draw=white,line width=1pt,fill=hilbert,text=white,
    font=\rmfamily\bfseries\fontsize{8}{9}\selectfont,inner sep=2.6pt},
  heading/.style={font=\rmfamily\bfseries\fontsize{11}{13}\selectfont},
  smallnote/.style={font=\rmfamily\fontsize{9}{11.5}\selectfont,text=muted},
]

\filldraw[fill=strongfill,draw=strongborder,rounded corners=5pt,line width=.65pt]
  (0,16.35) rectangle (23.4,20.05);
\filldraw[fill=weakfill,draw=panelborder,rounded corners=5pt,line width=.65pt]
  (0,1.35) rectangle (23.4,14.55);

\node[anchor=west,font=\rmfamily\bfseries\fontsize{18}{21}\selectfont]
  at (0,21.55) {How the ten metrics are related};
\node[anchor=west,text=muted,font=\rmfamily\fontsize{10}{12}\selectfont]
  at (0,20.85) {All distances have diameter one on $\mathcal P(\mathbb S^{d-1})$, with $d\geq2$, $1\leq p<\infty$, and fixed $\tau>0$.};

\node[heading,anchor=west] at (.4,19.55) {TOTAL-VARIATION TOPOLOGY};
\node[smallnote,anchor=east] at (23,19.55) {Mutually singular measures have distance $1$.};
\node[massmetric] (H) at (3.35,17.95)
  {\textbf{Hellinger}\\[5pt]
   $\displaystyle H=\sqrt{1-\int\sqrt{fg}\,dr}$};
\node[massmetric,text width=3.5cm] (V) at (11.7,17.95)
  {\textbf{Total variation}\\[5pt]
   $\displaystyle V=\frac12\int|f-g|\,dr$};
\node[massmetric] (J) at (20.05,17.95)
  {\textbf{Jensen-Shannon}\\[5pt]
   $\displaystyle J=\sqrt{\mathrm{JS}/\log 2}$};
\node[badge] at ($(H.north east)+(-.04,.02)$) {H};
\node[badge] at ($(J.north east)+(-.04,.02)$) {H};
\draw[boundlink] (H.east) -- (V.west)
  node[midway,above=4pt,edgeword] {$H^2\leq V\leq\sqrt2\,H$};
\draw[boundlink] (V.east) -- (J.west)
  node[midway,above=4pt,edgeword] {$\dfrac{V}{\sqrt{2\log2}}\leq J\leq\sqrt V$};

\draw[boundedge,line width=1.1pt] (V.south) -- (11.7,14.65);
\node[edgeword,fill=white,inner xsep=12pt,inner ysep=4pt] at (11.7,15.45)
  {$D_{p,\mathrm g},\,D_{p,\mathrm c}\leq V^{1/p},\qquad D_\tau\leq V,\qquad D_{\mathrm E}\leq V$\\[3pt]
   \textcolor{muted}{TV convergence implies weak convergence. The converse fails.}};

\node[heading,anchor=west] at (.4,14.03) {WEAK TOPOLOGY};
\node[smallnote,anchor=east] at (23,14.03) {All seven exact metrics below induce the same convergence.};

\node[metric] (G) at (3.35,12.05)
  {\textbf{Geodesic Wasserstein}\\[5pt]
   $\displaystyle D_{p,\mathrm g}=W_{p,\mathrm g}/\pi$};
\node[metric] (C) at (11.7,12.05)
  {\textbf{Chordal Wasserstein}\\[5pt]
   $\displaystyle D_{p,\mathrm c}=W_{p,\mathrm c}/2$};
\node[metric] (S) at (20.05,12.05)
  {\textbf{Sliced Wasserstein}\\[5pt]
   $\displaystyle D_{\mathrm{SW}}=\frac{\sqrt d}{2}\,\mathrm{SW}_2$};
\node[badge] at ($(S.north east)+(-.04,.02)$) {H};
\draw[boundlink] (G.east) -- (C.west)
  node[midway,above=4pt,edgeword] {$D_{p,\mathrm g}\leq D_{p,\mathrm c}$\\[2pt]
    $D_{p,\mathrm c}\leq\frac\pi2D_{p,\mathrm g}$};
\draw[boundedge] (C.east) -- (S.west)
  node[midway,above=5pt,edgeword] {$D_{\mathrm{SW}}\leq D_{2,\mathrm c}$};

\node[metric] (O) at (11.7,8.65)
  {\textbf{OT angular}\\[5pt]
   $\displaystyle A_{\mathrm{OT}}=\frac{\arccos(1-2D_{2,\mathrm c}^{\,2})}{\pi}$};
\node[metric] (A) at (20.05,8.65)
  {\textbf{Sliced angular}\\[5pt]
   $\displaystyle A_{\mathrm{SW}}=\frac{\arccos(1-2D_{\mathrm{SW}}^{\,2})}{\pi}$};
\draw[exactedge] (C.south) -- (O.north)
  node[midway,left=5pt,exactword] {$A_{\mathrm{OT}}=F(D_{2,\mathrm c})$};
\draw[exactedge] (S.south) -- (A.north)
  node[midway,left=5pt,exactword] {$A_{\mathrm{SW}}=F(D_{\mathrm{SW}})$};
\draw[boundedge] (O.east) -- (A.west)
  node[midway,above=5pt,edgeword] {$A_{\mathrm{SW}}\leq A_{\mathrm{OT}}$};
\draw[boundedge,rounded corners=5pt] (G.south) -- (3.35,8.65) -- (O.west)
  node[pos=.72,above=5pt,edgeword] {$A_{\mathrm{OT}}\leq D_{2,\mathrm g}$};
\node[smallnote,align=center] at (11.7,7.23)
  {$\displaystyle F(t)=\frac2\pi\arcsin t$.\quad Each dashed chord-to-angle pair has identical pairwise distance rankings.};

\draw[panelborder,line width=.55pt] (.5,6.70) -- (22.9,6.70);
\node[metric,text width=5.1cm] (M) at (4.05,4.67)
  {\textbf{Gaussian MMD}\\[5pt]
   $\displaystyle D_\tau=\frac{\mathrm{MMD}_\tau}{\sqrt{2(1-e^{-2\tau})}}$};
\node[metric,text width=5.1cm] (E) at (19.35,4.67)
  {\textbf{Chordal energy}\\[5pt]
   $\displaystyle D_{\mathrm E}=\frac12\sqrt{\mathcal E_{\mathrm c}}$};
\node[badge] at ($(M.north east)+(-.04,.02)$) {H};
\node[badge] at ($(E.north east)+(-.04,.02)$) {H};
\draw[mixedge] (M.east) -- (E.west)
  node[midway,above=5pt,exactword] {exact scale-integral identity};
\node[exactword] at (11.7,3.05)
  {$\displaystyle \mathcal E_{\mathrm c}
    =\frac1{\sqrt{2\pi}}\int_0^\infty \mathrm{MMD}_\tau^{\,2}\,\tau^{-3/2}\,d\tau$};
\node[smallnote] at (11.7,1.97)
  {$k_\tau(x,y)=e^{-\tau\|x-y\|^2/2}$.\quad Slicing uses the full uniform probability integral over directions.};

\draw[boundlink] (.25,.52) -- (1.35,.52);
\node[smallnote,anchor=west] at (1.55,.52) {solid: comparison bounds};
\draw[mixedge] (7.5,.52) -- (8.65,.52);
\node[smallnote,anchor=west] at (8.85,.52) {dashed: exact identities};
\node[badge] at (16.3,.52) {H};
\node[smallnote,anchor=west] at (16.6,.52) {Hilbertian metric};
\node[smallnote,anchor=west] at (.25,-.20)
  {Unmarked nodes are not Hilbertian on the full measure space. All distances use the same normalizations as the summary table.};
\node[smallnote,anchor=west] at (.25,-.75)
  {Here $r=\mu+\nu$, $f=d\mu/dr$, and $g=d\nu/dr$. The divergence $\mathrm{JS}$ uses natural logarithms. Selected connections are shown.};
\end{tikzpicture}%
  }%
}
\caption[Relationships among the ten spherical probability metrics]{Relationships among the ten metrics in Table~\ref{tab:main}. Solid blue connections give bounds, while dashed green connections give exact formulas. Each shaded region groups distances with the same convergent sequences, with total-variation convergence above and weak convergence below. An \textbf{H} badge marks an exact Hilbert norm distance. The integral formula uses $\MMD_\tau$ before normalization, whereas the metric nodes use distances scaled to have maximum one.}
\label{fig:metric-relationships}
\end{figure}
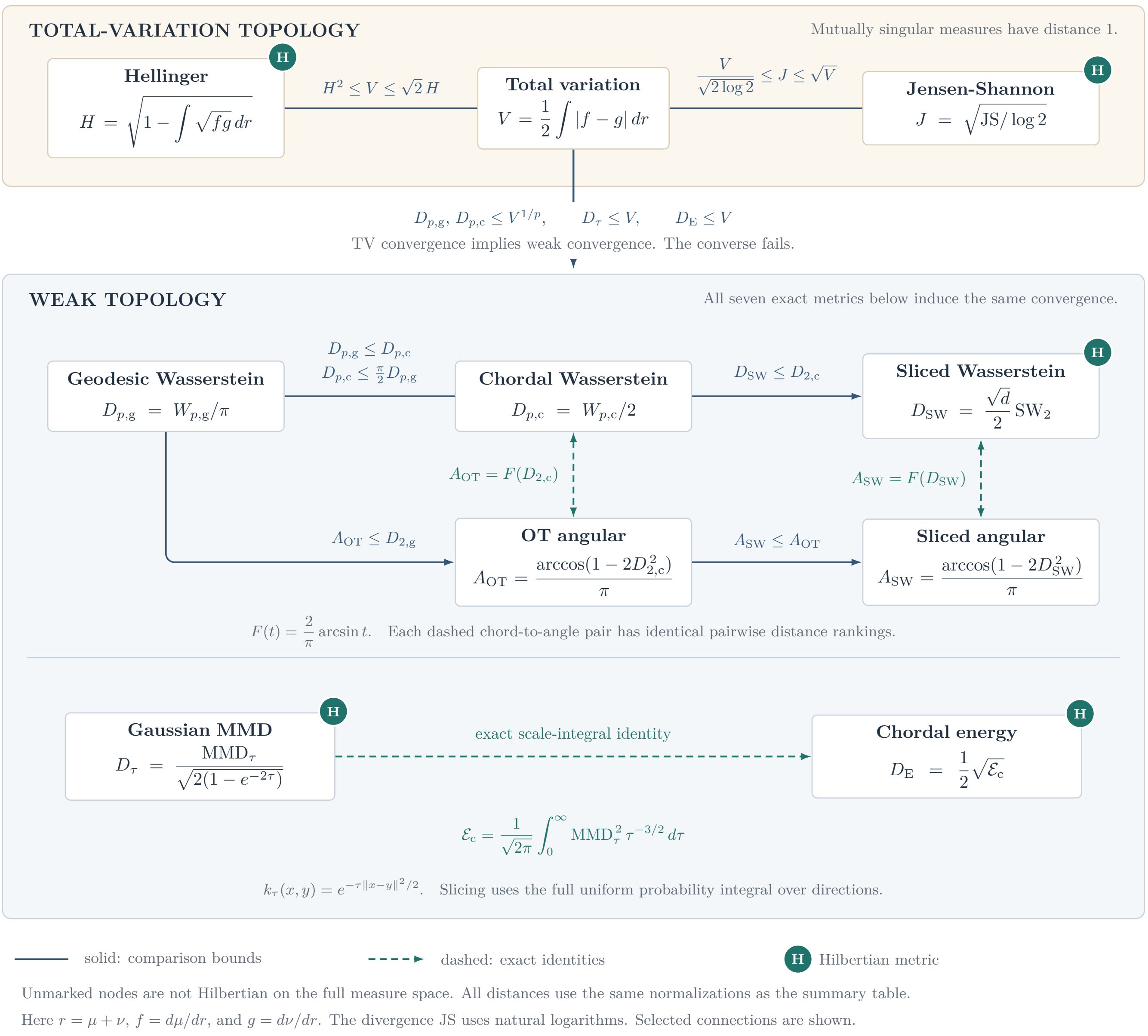

Table~\ref{tab:main} compares the properties of all ten metrics, while Figure~\ref{fig:metric-relationships} shows their inequalities and exact relationships. Throughout, an \emph{exact} distance uses the mathematical definition before an integral or optimization is replaced by a numerical approximation. The following definitions explain the properties compared in both displays.

\paragraph{Metric or pseudometric.}
A \emph{metric} is a finite, nonnegative distance. It is symmetric, gives zero exactly when its two arguments agree, and satisfies the triangle inequality
\[
 D(\mu,\lambda)\leq D(\mu,\nu)+D(\nu,\lambda).
\]
The requirement that only equal arguments have zero distance is called \emph{separation}. A \emph{pseudometric} has all the other properties, but may give zero to two different measures. A function that fails the triangle inequality is not even a pseudometric.

\paragraph{Hilbertian.}
A Hilbert space is an inner-product space in which every Cauchy sequence converges. Examples include Euclidean space and spaces of square-integrable functions. A metric is \emph{Hilbertian} when each measure can be represented by a vector $\Phi(\mu)$ in such a space so that
\[
 D(\mu,\nu)=\norm{\Phi(\mu)-\Phi(\nu)}.
\]
The representation must be \emph{injective}, or one-to-one, so different measures give different vectors. Thus the column asks whether the distance is exactly a norm distance between vector representations, possibly in infinitely many dimensions.

\paragraph{Weak convergence.}
An affirmative entry in the weak column means that convergence in the distance is exactly weak convergence,
\begin{equation*}
 D(\mu_n,\mu)\longrightarrow0\quad\Longleftrightarrow\quad\mu_n\Rightarrow\mu.
\end{equation*}
We express this property by saying that the distance \emph{metrizes weak convergence}. This property determines whether weakly converging approximations must become close in the chosen distance. A negative entry means that the distance describes a different notion of convergence, without implying that the distance is invalid.

\paragraph{Mutually singular measures.}
The last column asks whether \emph{every} mutually singular pair has distance one. A negative entry allows some such pairs to have distance one. Likewise, a negative entry in the Hilbertian column concerns the whole space $\Prob$, although a smaller class of measures may have a Hilbert representation.

\begin{table}[!t]
\centering
\small
\setlength{\tabcolsep}{4pt}
\renewcommand{\arraystretch}{1.55}
\begin{tabularx}{\textwidth}{@{}>{\raggedright\arraybackslash}p{3.0cm} Y c c c@{}}
\toprule
\textbf{Metric} & \textbf{Normalized distance} & \textbf{Weak} & \textbf{Hilbertian} & \textbf{Singular}\\
\midrule
Geodesic\newline Wasserstein & $\displaystyle D_{p,\mathrm g}=\frac{W_{p,\mathrm g}}{\pi}$ & Yes & No & No\\
Chordal\newline Wasserstein & $\displaystyle D_{p,\mathrm c}=\frac{W_{p,\mathrm c}}{2}$ & Yes & No & No\\
OT angular & $\displaystyle A_{\mathrm{OT}}=\frac{\arccos(1-W_{2,\mathrm c}^{\,2}/2)}{\pi}$ & Yes & No & No\\
Sliced Wasserstein & $\displaystyle D_{\mathrm{SW}}=\frac{\sqrt d}{2}\operatorname{SW}_2$ & Yes & Yes & No\\
Sliced angular & $\displaystyle A_{\mathrm{SW}}=\frac{\arccos(1-d\operatorname{SW}_2^{\,2}/2)}{\pi}$ & Yes & No & No\\
Gaussian MMD & $\displaystyle D_\tau=\frac{\MMD_\tau}{\sqrt{2(1-e^{-2\tau})}}$ & Yes & Yes & No\\
Chordal energy & $\displaystyle D_{\mathrm E}=\frac12\sqrt{\mathcal E_{\mathrm c}}$ & Yes & Yes & No\\
Total variation & $\displaystyle V=\frac12\int|f-g|\dd r$ & No & No & Yes\\
Hellinger & $\displaystyle H=\left[\frac12\int(\sqrt f-\sqrt g)^2\dd r\right]^{1/2}$ & No & Yes & Yes\\
Jensen-Shannon & $\displaystyle J=\sqrt{\frac{\JS}{\log2}}$ & No & Yes & Yes\\
\bottomrule
\end{tabularx}
\caption{Properties of the ten metrics on $\Prob$. Each has sharp range $[0,1]$, where $1\leq p<\infty$ and $\tau>0$ is fixed. The weak column asks whether convergence in the distance is exactly weak convergence. The Hilbertian column asks for an exact representation as a norm distance. The singular column asks whether every mutually singular pair has distance one. The sliced constructions use the full uniform probability average over directions.}\label{tab:main}
\end{table}

\begin{samepage}
\paragraph{Definitions used in the table.}
For Gaussian maximum mean discrepancy (MMD), use the similarity function $k_\tau(x,y)=e^{-\tau(1-x^\top y)}$, with $\tau>0$. Let $X,X'$ be independent draws from $\mu$, and $Y,Y'$ independent draws from $\nu$, with all four draws independent. Averaging over these draws gives the definitions
\begin{align}
 \MMD_\tau^2(\mu,\nu)
 &=\E k_\tau(X,X')+\E k_\tau(Y,Y')-2\E k_\tau(X,Y),\label{eq:mmddef}\\
 \Energy(\mu,\nu)
 &=2\E\norm{X-Y}-\E\norm{X-X'}-\E\norm{Y-Y'},\label{eq:energydef}\\
 \JS(\mu,\nu)
 &=\tfrac12\KL(\mu\Vert m)+\tfrac12\KL(\nu\Vert m),\qquad m=\tfrac12(\mu+\nu).\label{eq:jsdef}
\end{align}
\end{samepage}
Here $\KL$ is the Kullback-Leibler divergence \citep{kullback_1951_InformationSufficiency}. For discrete probabilities it is $\KL(a\Vert b)=\sum_i a_i\log(a_i/b_i)$. For measures it is $\int\log(d\mu/d\nu)\,d\mu$ when $\mu$ has a density relative to $\nu$, and is infinite otherwise. Terms with zero numerator contribute zero. Transport and sliced transport are defined in Sections~\ref{sec:wasserstein} and~\ref{sec:sliced}.

All ten distances apply to arbitrary probability measures, including empirical measures. Rotating or reflecting both measures in the same way leaves each distance unchanged. Point masses at opposite points of the sphere have distance one under every metric considered here. The later sections distinguish evaluating a distance between empirical measures, approximating it numerically, and estimating a distance between unknown population distributions.

\subsection{Two facts about vector representations}
Two facts about Hilbert spaces will be used repeatedly. The first characterizes exact Hilbert-space representations of distances. The second establishes that angles between unit vectors define a metric, even though they need not define a norm distance. Readers can first focus on the statements and the examples that use them.

A real symmetric function $K(z,z')$ is a \emph{positive semidefinite kernel} if every finite matrix $[K(z_i,z_j)]$ is positive semidefinite. Equivalently, $\sum_{i,j}b_i b_jK(z_i,z_j)\geq0$ for all real coefficients $b_i$. Such a matrix is also called a Gram matrix, because its entries can be written as inner products of vectors. The following classical result connects this property to distances \citep{schoenberg_1938_MetricSpacesPositive}.

\begin{lemma}[Hilbertian criterion]\label{lem:cnd}
A metric $D$ on a set $\mathcal X$ is Hilbertian if and only if
\begin{equation}\label{eq:cnd}
 \sum_{i,j=1}^N a_i a_jD(z_i,z_j)^2\leq0
 \quad\text{whenever }\sum_{i=1}^N a_i=0.
\end{equation}
Equivalently, $e^{-tD(z,z')^2}$ is a positive semidefinite kernel for every $t>0$. Condition~\eqref{eq:cnd} is called \emph{conditional negative definiteness of the squared distance}. For a useful test, consider four objects with distance $a$ between adjacent pairs in a cycle and distance $b$ between opposite pairs. If $b>\sqrt2a$, no Hilbert representation exists.
\end{lemma}

The criterion concerns distances measured by Hilbert norms. Angles give another comparison of unit vectors, and the next result relates them to chord lengths.

\begin{lemma}[Angles of unit Hilbert vectors]\label{lem:angle}
On the unit sphere of a real Hilbert space, $\alpha(u,v)=\arccos\ip uv$ is a metric. It is related to the chord by
\begin{equation}\label{eq:anglechord}
 \frac{\alpha(u,v)}{\pi}=\frac2\pi\arcsin\left(\frac{\norm{u-v}}2\right).
\end{equation}
An angular representation does not imply that the angular metric is Hilbertian.
\end{lemma}

\section{Geodesic and chordal Wasserstein metrics}\label{sec:wasserstein}
Wasserstein distance asks how far probability mass must move to turn one distribution into the other. A coupling describes a transport plan by pairing starting and ending points while preserving the required amounts of mass. The \emph{ground metric} $\rho$ measures the distance between individual points. For $1\leq p<\infty$, define
\begin{equation*}
 W_{p,\rho}(\mu,\nu)=
 \left[\inf_{\gamma\in\Pi(\mu,\nu)}\int \rho(x,y)^p\dd\gamma(x,y)\right]^{1/p}.
\end{equation*}
The integral is the average $p$th power of the distance moved under a transport plan. The infimum chooses the smallest possible cost, and the final $p$th root is essential to the metric. For $p=1$, this is the least average distance moved. For $p=2$, it is the square root of the least average squared distance. Choosing $\rho_{\mathrm g}$ measures movement along the sphere, while choosing $\rho_{\mathrm c}$ measures straight-line displacement. These give the standard distances $W_{p,\mathrm g}$ and $W_{p,\mathrm c}$ \citep{peyre_2019_ComputationalOptimalTransport}.

\begin{theorem}[Metric properties, diameter, and topology]\label{thm:Wp}
For either ground metric, $W_{p,\rho}$ is a metric on $\Prob$. If $B=\pi$ for $\rho_{\mathrm g}$ and $B=2$ for $\rho_{\mathrm c}$, then $W_{p,\rho}/B$ has diameter one and metrizes weak convergence. Distance one occurs exactly for antipodal point masses.
\end{theorem}

\subsection{Connections between ground geometries and transport orders}
The ground geometries give the transport comparisons
\begin{equation*}
 \frac2\pi W_{p,\mathrm g}\leq W_{p,\mathrm c}\leq W_{p,\mathrm g},
 \qquad
 D_{p,\mathrm g}\leq D_{p,\mathrm c}\leq\frac\pi2D_{p,\mathrm g}.
\end{equation*}
Each normalized distance is bounded by a fixed multiple of the other, a property called \emph{bi-Lipschitz equivalence}. The two distances therefore describe the same convergence, but they need not rank pairs of measures in the same order.

For a fixed ground metric and $1\leq p<q<\infty$, write $D_p=W_{p,\rho}/B$. Then
\begin{equation*}
 D_p\leq D_q\leq D_p^{p/q}.
\end{equation*}
Changing $p$ preserves convergence to zero, but changes how the distance responds when a small amount of mass moves a long way.

\subsection{Why neither Wasserstein family is Hilbertian}
Neither Wasserstein family is Hilbertian on the full space of measures. Four point masses at successive quarter-turns of a great circle give a counterexample for geodesic Wasserstein.

For chordal Wasserstein, consider the antipodal pairs
\[
 u_j=(\cos(j\pi/4),\sin(j\pi/4),0,\ldots,0),\qquad
 \mu_j=\tfrac12(\delta_{u_j}+\delta_{-u_j}),\quad j=0,1,2,3.
\]
For every finite $p\geq1$, their distances satisfy
\begin{equation}\label{eq:antipodalWp}
 W_{p,\mathrm c}(\mu_i,\mu_j)^2=2-2|u_i^\top u_j|.
\end{equation}
These four measures rule out a Hilbert representation for every stated transport order.

\paragraph{Connection to probability-mass distances.}
Both Wasserstein families respect small geometric movement, even when supports are disjoint. In particular, $D_{p,\rho}(\delta_x,\delta_y)=\rho(x,y)/B\to0$ as $y\to x$. This contrasts with total variation, Hellinger, and Jensen-Shannon distance. The exact cosine connection for chordal transport at $p=2$ is the subject of the next section.

\section{Optimal cosine matching and the OT angular metric}\label{sec:otangle}
Cosine similarity compares two individual directions. To compare two distributions, choose the coupling that gives the largest average cosine similarity. This gives the optimal-transport (OT) cosine matching score

\begin{equation}\label{eq:cot}
 \Cot(\mu,\nu)=\max_{\gamma\in\Pi(\mu,\nu)}\int x^\top y\dd\gamma(x,y).
\end{equation}
The matching score has the exact relation
\begin{equation}\label{eq:cotidentity}
 \Cot=1-\tfrac12W_{2,\mathrm c}^2=1-2D_{2,\mathrm c}^2.
\end{equation}
Thus maximizing average cosine is exactly the same optimization as minimizing squared chordal transport cost. The score $\Cot$ lies in $[-1,1]$. Taking its inverse cosine gives the angular distance below. This angle also appears in the study of Wasserstein geometry by \citet{takatsu_2012_CONESTRUCTUREL2}.

\begin{theorem}[The OT angle]\label{thm:otangle}
The function
\begin{equation}\label{eq:aot}
 \Aot(\mu,\nu)=\frac1\pi\arccos\Cot(\mu,\nu)
 =\frac2\pi\arcsin D_{2,\mathrm c}(\mu,\nu)
\end{equation}
is a metric of diameter one, metrizes weak convergence, and is not Hilbertian. It reduces to $\rho_{\mathrm g}(x,y)/\pi$ on point masses. Moreover,
\begin{equation}\label{eq:aotbounds}
 \frac2\pi D_{2,\mathrm c}\leq\Aot\leq D_{2,\mathrm c},
 \qquad \Aot\leq D_{2,\mathrm g}.
\end{equation}
\end{theorem}

The chordal and angular distances in~\eqref{eq:aot} rank neighbors in exactly the same order, including ties, because their relationship is strictly increasing. They can still give different results in procedures that add or square distance values. More generally, an increasing transformation of a metric need not preserve the triangle inequality.

A simple mixture separates the OT angle from geodesic Wasserstein. For
\[
 \mu=\delta_x,\qquad \nu=\tfrac34\delta_x+\tfrac14\delta_{-x},
\]
we have $\Aot=1/3$, $D_{1,\mathrm g}=1/4$, and $D_{2,\mathrm g}=1/2$.

The cosine $\Cot$ does not always give a positive semidefinite similarity matrix. The antipodal measures in~\eqref{eq:antipodalWp} provide a counterexample, even after rescaling the scores to $(1+\Cot)/2$. In contrast, the sliced cosine in the next section always gives a positive semidefinite Gram matrix.

Finally, independent rather than optimal matching gives
\[
 \iint x^\top y\dd\mu(x)\dd\nu(y)
 =\left(\int x\dd\mu\right)^\top\left(\int y\dd\nu\right),
\]
so independent matching retains only the first moments of the two distributions. Optimal matching need not share this loss of information.

\section{Sliced Wasserstein and its angle}\label{sec:sliced}
Instead of matching points in the full space, project them onto a line. For a unit direction $\theta$, a point $x$ has coordinate $\theta^\top x$ on that line. Write $\mu_\theta=(x\mapsto\theta^\top x)_\#\mu$ for the distribution of these coordinates, which lie in $[-1,1]$. Its \emph{quantile function} gives the value at probability level $t$,

\[
 Q_\mu(\theta,t)=\inf\{a:\mu_\theta(({-\infty},a])\geq t\},\qquad 0<t<1.
\]
For an equally weighted finite sample, quantiles are obtained by sorting the projected values. One-dimensional transport then pairs equal probability levels in the two quantile functions. The quantiles are measurable jointly in $(\theta,t)$. Also $|Q_\mu|\leq1$, so the following integrals exist. Averaging the one-dimensional squared transport distances over all directions gives
\begin{equation*}
 \SW(\mu,\nu)^2
 =\int_{\Sph}W_2(\mu_\theta,\nu_\theta)^2\dd\sigma(\theta)
 =\int_{\Sph}\int_0^1|Q_\mu(\theta,t)-Q_\nu(\theta,t)|^2\dd t\dd\sigma(\theta).
\end{equation*}
This construction projects onto straight lines in $\R^d$, not onto great circles of the sphere. The quantile transport formula is treated in \citet[Chapter~2]{santambrogio_2015_OptimalTransportApplied}. It is also called \emph{parallel slicing}. See \citet{quellmalz_2024_ParallellySlicedOptimal} for measures on spheres and \citet{kolouri_2016_SlicedWassersteinKernels} for the quantile representation and associated kernels.

The quantile formula also applies to distributions with point masses. The next result turns a distribution into a unit vector of quantile values. Distances between these vectors give sliced Wasserstein, and angles between them give its angular version. Here $\mathrm{Leb}$ denotes Lebesgue measure on $(0,1)$.

\begin{theorem}[Unit-norm quantile representation]\label{thm:sw}
Define $\Phi_\mu(\theta,t)=\sqrt d\,Q_\mu(\theta,t)$ in
$\mathcal H=L^2(\Sph\times(0,1),\sigma\otimes\mathrm{Leb})$. Then $\norm{\Phi_\mu}=1$, the map $\mu\mapsto\Phi_\mu$ is injective, and
\begin{align}
 \Dsw(\mu,\nu)&=\frac12\norm{\Phi_\mu-\Phi_\nu}=\frac{\sqrt d}{2}\SW(\mu,\nu),\label{eq:swembed}\\
 \Csw(\mu,\nu)&=\ip{\Phi_\mu}{\Phi_\nu}=1-\frac d2\SW(\mu,\nu)^2,\label{eq:csw}\\
 \Asw(\mu,\nu)&=\frac1\pi\arccos\Csw(\mu,\nu)
 =\frac2\pi\arcsin\Dsw(\mu,\nu).\label{eq:asw}
\end{align}
Both distances are metrics of diameter one and metrize weak convergence. The chordal distance $\Dsw$ is Hilbertian, whereas the angular distance $\Asw$ is not. The cosine $\Csw$ is a positive semidefinite kernel.
\end{theorem}

The quantile representation also links sliced distance to ordinary means and full transport. The following bounds make these comparisons precise.

\begin{proposition}\label{prop:swbounds}
Writing $m_\mu=\int x\dd\mu(x)$, we have
\begin{equation}\label{eq:swbounds}
 \frac12\norm{m_\mu-m_\nu}\leq\Dsw\leq D_{2,\mathrm c},
 \qquad \Asw\leq\Aot,
 \qquad \Csw\geq\Cot.
\end{equation}
\end{proposition}

Each projection may use a different optimal matching. These matchings need not come from one common matching of the original points, so the upper bound can be strict. A more accurate uniform average approaches the sliced distance, which can differ from full Wasserstein distance. As Section~\ref{sec:approx} explains, the selected directions must approximate the uniform average, so increasing their number alone does not suffice. For example, on the circle let
$\mu=(\delta_{e_1}+\delta_{-e_1})/2$ and $\nu=(\delta_{e_2}+\delta_{-e_2})/2$.
For this pair, $\Cot=0$ while $\Csw=2/\pi>0$. Thus full and sliced matching agree for individual point masses but can disagree for distributions. Sliced Wasserstein and its angle reach one only for point masses at opposite points.

\section{Gaussian maximum mean discrepancy}\label{sec:gaussian}
Kernel methods represent each point $x$ by a vector $\phi_k(x)$ and compare averages of these vectors. The vectors may have infinitely many coordinates, but a kernel gives their inner products directly, $\langle\phi_k(x),\phi_k(y)\rangle=k(x,y)$. Every positive semidefinite kernel has such a Hilbert-space representation \citep{aronszajn_1950_TheoryReproducingKernels}. The average $m_k(\mu)=\int\phi_k(x)\dd\mu(x)$ is called the \emph{kernel mean embedding}, reviewed in \citet{muandet_2017_KernelMeanEmbedding}. For a bounded continuous kernel on the compact sphere, the feature map is bounded and continuous, so this vector-valued integral exists. Maximum mean discrepancy measures the distance between the two averages,
\begin{equation*}
 \MMD_k(\mu,\nu)=\norm{m_k(\mu)-m_k(\nu)}.
\end{equation*}
Unlike an ordinary mean vector, a kernel mean can retain enough information to identify the whole distribution. A norm distance gives symmetry and the triangle inequality. It is a metric on distributions when $m_k(\mu)=m_k(\nu)$ implies $\mu=\nu$. A kernel with this property is called \emph{characteristic} \citep{sriperumbudur_2010_HilbertSpaceEmbeddings}, while a noncharacteristic kernel gives only a pseudometric. The Gaussian kernel below is characteristic, and its parameter $\tau$ controls how quickly pointwise similarity decreases with distance.

\begin{samepage}
\begin{theorem}[Gaussian MMD]\label{thm:mmd}
For fixed $\tau>0$, the kernel $k_\tau(x,y)=e^{-\tau\norm{x-y}^2/2}$ is characteristic, and
\[
 D_\tau(\mu,\nu)=\frac{\MMD_\tau(\mu,\nu)}{\sqrt{2(1-e^{-2\tau})}}
\]
is a Hilbertian metric of diameter one that metrizes weak convergence. Its diameter is attained only by antipodal point masses.
\end{theorem}
\end{samepage}

\subsection{Connections to transport, total variation, and first moments}
Gaussian MMD obeys the transport bound
\begin{equation*}
 D_\tau\leq
 \sqrt{\frac{2\tau}{1-e^{-2\tau}}}\,D_{1,\mathrm c}.
\end{equation*}
It also satisfies $D_\tau\leq\TV$. The first bound controls the response to geometric displacement, while the second controls the response to changes in probability mass.

As $\tau$ tends to zero, the kernel becomes less sensitive to locations. The limiting distance retains only the ordinary mean vectors, although every fixed $\tau>0$ distinguishes full distributions. The expansion and limit are
\begin{equation}\label{eq:mmdsmalltau}
 \MMD_\tau^2=\tau\norm{m_\mu-m_\nu}^2+O(\tau^2),
 \qquad
 D_\tau\longrightarrow\frac12\norm{m_\mu-m_\nu}.
\end{equation}
The last expression can vanish for different measures, as the two perpendicular antipodal distributions in Section~\ref{sec:sliced} both have mean zero. The table's metric claim therefore requires a fixed positive $\tau$. The limit illustrates the information lost by comparisons based only on means.

\subsection{A related cosine between kernel means}
A related construction compares the normalized kernel-mean vectors. It lies outside the ten metrics in Table~\ref{tab:main} and uses
\[
 K_\tau(\mu,\nu)=\iint k_\tau\dd\mu\dd\nu,
 \qquad
 C_\tau(\mu,\nu)=\frac{K_\tau(\mu,\nu)}{\sqrt{K_\tau(\mu,\mu)K_\tau(\nu,\nu)}}.
\]
The denominator is positive, and $e^{-2\tau}\leq C_\tau\leq1$. This is ordinary cosine similarity between the two kernel-mean vectors. For the Gaussian kernel, normalization preserves the distinction between measures. Consequently both
\begin{equation*}
 \sqrt{\frac{1-C_\tau}{1-e^{-2\tau}}},
 \qquad
 \frac{\arccos C_\tau}{\arccos(e^{-2\tau})}
\end{equation*}
are metrics with maximum value one. As with sliced Wasserstein and its angle, they measure chords and angles between unit vectors, here obtained from kernel means. The cosine $C_\tau$ always gives a positive semidefinite matrix, a property that $\Cot$ lacks.

\section{Chordal energy and its connection to Gaussian MMD}\label{sec:energy}
Energy distance compares average distances between two distributions with average distances within each one. In~\eqref{eq:energydef}, the first term measures between-distribution distances, and the two subtracted terms account for the spread within the distributions. These averages require no optimal matching. For the foundations of energy statistics, see \citet{szekely_2013_EnergyStatisticsClass}, and for its connection to kernel distances, see \citet{sejdinovic_2013_EquivalenceDistancebasedRKHSbased}. Both the kernel and its normalization take simple forms on the sphere.

\begin{theorem}[Energy metric]\label{thm:energy}
The function $\DE=\tfrac12\sqrt{\Energy}$ is a Hilbertian metric of diameter one that metrizes weak convergence. It has the representations
\begin{align}
 k_{\mathrm E}(x,y)&=1-\tfrac12\norm{x-y},\qquad
 \Energy=2\MMD_{k_{\mathrm E}}^2,\label{eq:energykernel}\\
 \Energy(\mu,\nu)&=\frac1{\sqrt{2\pi}}
 \int_0^\infty\MMD_\tau(\mu,\nu)^2\tau^{-3/2}\dd\tau.\label{eq:energymixture}
\end{align}
The kernel in~\eqref{eq:energykernel} is positive semidefinite and characteristic on the sphere. Moreover,
\begin{equation}\label{eq:energybounds}
 \DE^2\leq D_{1,\mathrm c},\qquad \DE\leq\TV.
\end{equation}
\end{theorem}

Energy uses averages of independent point-pair distances without computing an optimal transport plan. The normalized chordal energy distance $\DE$ is exactly a scaled kernel norm distance. Equation~\eqref{eq:energymixture} expresses the unrooted energy $\Energy$ as a mixture of squared Gaussian MMD values over all positive $\tau$. Chordal energy therefore requires no choice of a single kernel scale. For point masses, $\DE(\delta_x,\delta_y)=\sqrt{\norm{x-y}/2}$, which already differs from normalized chordal Wasserstein.

The choice of ground distance matters here. If one replaces $\norm{x-y}$ by $\norm{x-y}^2$ in the energy expression, expansion gives
\[
 2\E\norm{X-Y}^2-\E\norm{X-X'}^2-\E\norm{Y-Y'}^2
 =2\norm{m_\mu-m_\nu}^2.
\]
Taking a square root leaves only a comparison of mean vectors. Up to a constant factor, this is the pseudometric in~\eqref{eq:mmdsmalltau}, which cannot distinguish all distributions.

Using geodesic distance in the energy formula can give zero to different measures. The perpendicular antipodal distributions in Section~\ref{sec:sliced} have zero geodesic energy. Geodesic Wasserstein has a separate metric proof based on optimal matching. Independent averages and optimal matching are different constructions, and each needs its own proof of separation.

The unrooted $\Energy$ also fails the triangle inequality. The midpoint mixture $\lambda=(\mu+\nu)/2$ gives a counterexample for every $\mu\ne\nu$, and the same construction applies to squared MMD.

\section{Total variation}\label{sec:tv}
Total variation measures how much the two distributions disagree about probabilities, without using the distances between point locations. Write their densities as $f,g$ relative to a common reference measure $r$. Then
\begin{equation}\label{eq:tv}
 V(\mu,\nu)=\TV(\mu,\nu)=\tfrac12\int|f-g|\dd r.
\end{equation}
The answer does not depend on the choice of $r$. Equivalently, $V=\sup_A|\mu(A)-\nu(A)|$, where the supremum is over measurable sets. Thus TV is the largest difference in probability assigned to the same event.

\begin{proposition}[Metric and singularity properties]\label{prop:tv}
Total variation is a diameter-one metric, and $V(\mu,\nu)=1$ if and only if $\mu\perp\nu$. It does not metrize weak convergence and is not Hilbertian on $\Prob$.
\end{proposition}

Total variation also bounds several distances that use geometry. The mass shared by two measures can remain in place, so only the residual mass needs to move. For a ground metric with diameter at most $B$, this gives
\begin{equation}\label{eq:tvtransport}
 \frac{W_{p,\rho}(\mu,\nu)}B\leq V(\mu,\nu)^{1/p}.
\end{equation}
For the linear kernel-mean representations, the corresponding bounds are
\begin{equation}\label{eq:tvkernel}
 D_\tau\leq V,\qquad \DE\leq V.
\end{equation}
These comparisons link transport, Gaussian MMD, and energy distance to total variation. The kernel argument requires linearity, so it does not apply to the nonlinear sliced quantile, square-root density, and Jensen-Shannon representations.

For empirical measures, collect all distinct observed locations, combine the weights of repeated points, and evaluate~\eqref{eq:tv} as a finite sum without smoothing. If the samples share no locations, TV equals one, however close their points may be. This remains a valid distance, but it answers a question about exact mass overlap rather than geometric closeness.

\section{Hellinger distance}\label{sec:hellinger}
Hellinger distance also compares probability mass, but it takes square roots of the densities before measuring their $L^2$ distance. For comparisons with other probability metrics and divergence-based distances, see \citet{gibbs_2002_ChoosingBoundingProbability,osterreicher_2003_NewClassMetric}. Define
\begin{equation}\label{eq:hellinger}
 H(\mu,\nu)^2=\tfrac12\int(\sqrt f-\sqrt g)^2\dd r
 =1-\int\sqrt{fg}\dd r.
\end{equation}
The integral $\int\sqrt{fg}\,dr$ is called the \emph{Hellinger affinity}. It measures overlap, equaling one for identical distributions and zero for mutually singular distributions. The answer is independent of the common reference measure.

\begin{theorem}[Metric, embedding, and comparisons]\label{thm:hellinger}
Hellinger distance is a Hilbertian diameter-one metric. It equals one exactly for mutually singular measures. For all pairs,
\begin{equation}\label{eq:Htv}
 H^2\leq V\leq H\sqrt{2-H^2}\leq\sqrt2H.
\end{equation}
Consequently it induces the total-variation topology, not the weak topology on $\Prob$.
\end{theorem}

\paragraph{Connections to the angular constructions.}
For measures with a common reference $r$, each square-root density $\sqrt f$ is a unit vector in $L^2(r)$, with inner product $1-H^2$ against $\sqrt g$. Hellinger is therefore a scaled chordal distance between unit vectors, like sliced Wasserstein. The shared vector geometry does not make the comparisons interchangeable. Square-root density distances measure exact mass overlap, whereas projected quantile distances also respond to the locations of the mass. The associated angle $\arccos(1-H^2)$ is another metric, outside the ten metrics summarized here.

Squaring destroys the triangle inequality here. For distinct points $x,y$ and $\lambda=(\delta_x+\delta_y)/2$,
\[
 H^2(\delta_x,\delta_y)=1,
 \qquad H^2(\delta_x,\lambda)+H^2(\lambda,\delta_y)=2-\sqrt2<1.
\]

\section{The Jensen-Shannon metric}\label{sec:js}
Jensen-Shannon divergence \citep{lin_1991_DivergenceMeasuresBased} compares each measure to their equal mixture $m=(\mu+\nu)/2$ using KL divergence, then averages the results as in~\eqref{eq:jsdef}. We use the convention $0\log0=0$. JS is finite even when the original measures have disjoint supports. Taking the square root gives a metric, as established by \citet{endres_2003_NewMetricProbability}, although the unrooted divergence fails the triangle inequality. Related metric divergences are studied by \citet{osterreicher_2003_NewClassMetric}. \citet{fuglede_2004_JensenShannonDivergenceHilbert} developed a Hilbert representation of the square-root distance. A scalar squared-distance identity gives an explicit construction of this representation.

\begin{theorem}[Metric, Hilbertian structure, and topology]\label{thm:js}
The function $J=\sqrt{\JS/\log2}$ is a Hilbertian diameter-one metric. It equals one exactly for mutually singular measures. Moreover,
\begin{equation}\label{eq:Jtv}
 \tfrac12 V^2\leq\JS\leq(\log2)V,
 \qquad
 \frac{V}{\sqrt{2\log2}}\leq J\leq\sqrt V.
\end{equation}
It induces the same topology as TV and Hellinger, which is strictly stronger than the weak topology.
\end{theorem}

\paragraph{Connections and nonmetric relatives.}
Like Gaussian MMD and normalized chordal energy, $J$ is a Hilbert norm distance, so $e^{-tJ^2}$ is a positive semidefinite kernel for every $t>0$. The distinctions are that $J$ is always one for disjoint supports and its vector representation is nonlinear in the measure. Squaring $J$ removes the triangle inequality, as seen by taking $\lambda=(\delta_x+\delta_y)/2$ with $x\ne y$.
\[
 \JS(\delta_x,\delta_y)=\log2,
 \qquad
 \JS(\delta_x,\lambda)+\JS(\lambda,\delta_y)=\tfrac32\log(4/3)<\log2.
\]
This counterexample explains why the square root belongs in the metric's definition.

\section{Connections across the table}\label{sec:connections}
The preceding results have consequences for rotation, maximum separation, small changes in a distribution, and convergence. Figure~\ref{fig:metric-relationships}, beside the summary table in Section~\ref{sec:table}, collects the relationships that we discuss here.

\subsection{Rotating both measures and reaching the maximum distance}
For every orthogonal matrix $U$ and every metric $D$ in Table~\ref{tab:main},
\[
 D(U_\#\mu,U_\#\nu)=D(\mu,\nu).
\]
This invariance requires a \emph{common} rotation of the two measures. It does not establish invariance under separate alignment.

A shared zero-to-one range does not give every metric the same interpretation of maximum separation. Full and sliced transport, their angles, Gaussian MMD, and chordal energy reach one only for point masses at opposite points. Total variation, Hellinger, and Jensen-Shannon distance reach one for every mutually singular pair. Disjoint supports alone therefore do not force maximum distance for the transport, kernel, or energy comparisons.

\subsection{A summary of quantitative comparisons}
The preceding bounds can be collected as follows.
\begin{align*}
 D_{p,\mathrm g}&\leq D_{p,\mathrm c}\leq\tfrac\pi2D_{p,\mathrm g},\\
 \Dsw&\leq D_{2,\mathrm c},\qquad
 \Asw\leq\Aot\leq D_{2,\mathrm g},\\
 \DE&\leq\min\{V,\sqrt{D_{1,\mathrm c}}\},\qquad
 D_\tau\leq\min\left\{V,\sqrt{\frac{2\tau}{1-e^{-2\tau}}}\,D_{1,\mathrm c}\right\},\\
 D_{p,\mathrm g},D_{p,\mathrm c}&\leq V^{1/p},\qquad
 H^2\leq V\leq\sqrt2H,\qquad
 \frac{V}{\sqrt{2\log2}}\leq J\leq\sqrt V.
\end{align*}
All constants use the normalizations in Table~\ref{tab:main}. A common numerical range would not yield these inequalities or imply identical rankings. The exact monotone chord-angle pairs in~\eqref{eq:aot} and~\eqref{eq:asw} do preserve rankings.

\subsection{How the distances respond to moving a little mass}
Start with all probability at one point $x$ and move a fraction $t$ to the opposite point, giving
\begin{equation*}
 \mu=\delta_x,\qquad\nu_t=(1-t)\delta_x+t\delta_{-x},\qquad0\leq t\leq1.
\end{equation*}
The resulting distances are
\begin{align}
 D_{p,\mathrm g}(\mu,\nu_t)=D_{p,\mathrm c}(\mu,\nu_t)&=t^{1/p},\notag
\\
 \Dsw(\mu,\nu_t)&=\sqrt t,\qquad
 \Aot(\mu,\nu_t)=\Asw(\mu,\nu_t)=\tfrac2\pi\arcsin\sqrt t,\label{eq:contam2}\\
 D_\tau(\mu,\nu_t)=\DE(\mu,\nu_t)=V(\mu,\nu_t)&=t,\label{eq:contam3}\\
 H(\mu,\nu_t)^2&=1-\sqrt{1-t},\notag
\\
 \JS(\mu,\nu_t)&=\tfrac12\left[-\log(1-t/2)
 +(1-t)\log\frac{1-t}{1-t/2}+t\log2\right].\label{eq:contam5}
\end{align}
The endpoint term $0\log0$ in~\eqref{eq:contam5} is understood as zero.

As $t\downarrow0$, both $H$ and $J$ behave like $\sqrt{t/2}$, where $\sim$ denotes a ratio tending to one. The normalized Gaussian MMD and chordal energy metrics, along with TV, instead equal $t$. These rates describe how the distances respond to this perturbation. They do not establish statistical power, which also depends on the sampling model and the test. A different perturbation moves a point mass by a tiny angle. Total variation, Hellinger, and Jensen-Shannon distance then remain one as long as the points differ. Moving a small \emph{amount} of mass and moving mass a small \emph{distance} therefore give different notions of a small change.

\subsection{Two groups of convergence properties}
Full and sliced transport, their angles, Gaussian MMD, and chordal energy describe weak convergence, although their numerical values differ. Total variation, Hellinger, and Jensen-Shannon distance describe stronger convergence. Three standard metric-space properties further distinguish these groups. A space is \emph{compact} if every sequence has a convergent subsequence. It is \emph{complete} if every Cauchy sequence converges, where the terms of a Cauchy sequence eventually become arbitrarily close to one another. A space is \emph{separable} if a countable collection of points can approximate every point arbitrarily well. Here the points are themselves probability measures.

\begin{proposition}\label{prop:topology}
Full and sliced Wasserstein distances, the OT and sliced angles, Gaussian MMD, and chordal energy induce the weak topology on $\Prob$. Their metric spaces are compact, complete, and separable. Total variation, Hellinger, and Jensen-Shannon distance induce the total-variation topology, and their spaces are complete but not separable. The latter topology is strictly stronger than the former.
\end{proposition}

The weak column describes whether empirical approximations that converge weakly must also approach the population in the chosen distance. For example, suppose $\mu$ gives zero probability to every individual point. Every finite empirical measure is then mutually singular to $\mu$, so its total variation, Hellinger, and Jensen-Shannon distances from $\mu$ equal one. If those empirical measures converge weakly to $\mu$, their full and sliced transport distances, corresponding angles, Gaussian MMD, and chordal energy distances tend to zero.

\section{Computing distances and understanding approximations}\label{sec:approx}
The table describes exact mathematical distances. Software may evaluate an exact distance between finite measures, approximate that distance numerically, or estimate a population quantity with a different formula. These tasks can produce quantities with different metric properties, as the examples below show.

\subsection{Exact finite formulas}
For $\mu=\sum_i a_i\delta_{x_i}$ and $\nu=\sum_j b_j\delta_{y_j}$, let $\Gamma_{ij}$ denote the mass moved from $x_i$ to $y_j$. The row sums are $a_i$, and the column sums are $b_j$. Both Wasserstein costs are therefore finite linear optimization problems.
\begin{equation*}
 W_{p,\rho}^p=\min_{\Gamma\geq0}\sum_{i,j}\Gamma_{ij}\rho(x_i,y_j)^p,
 \qquad \Gamma\1=a,\quad\Gamma^\top\1=b.
\end{equation*}
Here $\1$ denotes a vector of ones of the appropriate length. With equal weights and equal sample sizes, an optimal plan can be chosen as a one-to-one matching.

For Gaussian MMD and energy, replace the expectations in~\eqref{eq:mmddef} and~\eqref{eq:energydef} by full weighted sums over point pairs, including self-comparisons. For TV, Hellinger, and JS, sum over all distinct locations in either sample, assigning zero weight wherever a measure has no point. These formulas give exact distances between empirical measures, including those with different counts or weights.

To calculate projected $W_2$, sort the projected points and form their cumulative weights. The quantile functions are constant between successive cumulative-weight values. Split $(0,1)$ at every breakpoint from either sample, then multiply each squared quantile difference by its interval length and sum. This gives the exact one-dimensional cost for samples of any size.

\subsection{Distances between representations}
An implementation can first replace each object by a summary $T(x)$, such as selected projections or a finite feature vector. Comparing these summaries using a metric defines the \emph{pullback} of that metric. Its separation property depends on whether the summary retains enough information to distinguish the original objects.

\begin{proposition}[A representation must distinguish the objects]\label{prop:pullback}
Let $T:\mathcal X\to\mathcal Y$, where $d_{\mathcal Y}$ is a metric. Then $d_T(x,x')=d_{\mathcal Y}(Tx,Tx')$ is a pseudometric. It is a metric on $\mathcal X$ if and only if $T$ is injective. The same criterion applies to finite-feature norm distances and noncharacteristic kernel mean embeddings.
\end{proposition}

A fixed finite vector of average features cannot distinguish \emph{all} probability measures on the sphere.

\subsection{Normalization and separation for finite sliced distances}
For directions $\theta_\ell$ and weights $w_\ell>0$ summing to one, define
\begin{equation*}
 \SWL(\mu,\nu)^2=\sum_{\ell=1}^Lw_\ell
 W_2(\mu_{\theta_\ell},\nu_{\theta_\ell})^2.
\end{equation*}
The square root gives a pseudometric. It is a metric on a chosen class exactly when the selected projections distinguish all measures in that class. Equivalently, equality of every selected projected distribution must imply equality of the original measures. Using finitely many projections does not, by itself, settle this question.

For any choice of unit directions, $\SWL\leq2$, so $\SWL/2$ is a bounded pseudometric. The sharper normalization in Table~\ref{tab:main} requires the directions to balance all coordinate axes through their second moments. This condition is called \emph{isotropy} and takes the form
\begin{equation}\label{eq:tightframe}
 \sum_{\ell=1}^L w_\ell\theta_\ell\theta_\ell^\top=\frac1dI_d.
\end{equation}
Then the representation $\Phi_\mu^L(\ell,t)=\sqrt{d w_\ell}\,Q_\mu(\theta_\ell,t)$ in $\bigoplus_{\ell=1}^L L^2(0,1)$ has unit norm, and
\[
 D_L=\frac{\sqrt d}{2}\SWL,\qquad
 A_L=\frac1\pi\arccos(1-d\SWL^2/2)
\]\nopagebreak
are bounded pseudometrics. They become metrics if the selected projections distinguish the measures under study. An orthonormal basis with equal weights satisfies~\eqref{eq:tightframe}, as does any weighted mixture of such equally weighted bases.

Balanced directions can still miss a difference between distributions. For example, the two coordinate projections on the circle cannot distinguish the pair
\[
 \mu=\tfrac12\delta_{(1,1)/\sqrt2}+\tfrac12\delta_{(-1,-1)/\sqrt2},\qquad
 \nu=\tfrac12\delta_{(1,-1)/\sqrt2}+\tfrac12\delta_{(-1,1)/\sqrt2}.
\]
Both coordinate distributions agree, although the measures differ. The directions satisfy~\eqref{eq:tightframe}, and the same example extends to higher dimensions.

Arbitrary directions can make the angular formula undefined over the reals. One direction $\theta$ and the pair $\delta_\theta,\delta_{-\theta}$ already provide an example. Clipping the argument to $[-1,1]$ does not establish a triangle inequality. Alternatively, one can normalize each nonzero representation to unit length and compare angles. This gives a pseudometric wherever the normalization is defined, but separation still needs to be checked.

\paragraph{Convergence of the finite directional average.}
Let $\lambda_L=\sum_\ell w_{L\ell}\delta_{\theta_{L\ell}}$ be the probability distribution of the selected directions. A sufficient condition for convergence to full slicing is $\lambda_L\Rightarrow\sigma$, meaning that the weighted directions approach the uniform distribution. Under this condition, $\SWL^2(\mu,\nu)\to\SW^2(\mu,\nu)$ for every fixed pair. Repeating the coordinate directions in the example above, however many times, never distinguishes that pair. Isotropy, identification, and convergence of the directional average are separate requirements.

\subsection{An empirical metric is not an unbiased statistical estimator}
A distance between observed empirical measures differs from an estimator designed to recover a population quantity on average over repeated samples. Consider two independent i.i.d. samples of sizes $n,m\geq2$. An unbiased estimator of population MMD squared omits self-comparisons and adjusts the denominators.
\[
 \widehat{\MMD^2}_{\mathrm u}
 =\frac{\sum_{i\ne i'}k(X_i,X_{i'})}{n(n-1)}
 +\frac{\sum_{j\ne j'}k(Y_j,Y_{j'})}{m(m-1)}
 -\frac{2\sum_{i,j}k(X_i,Y_j)}{nm}.
\]
Its expectation equals the population MMD squared, but its value on a particular dataset can be negative \citep{gretton_2012_KernelTwoSampleTest}. For example, let both samples contain the same two points $u,v$, with $k(u,u)=k(v,v)=1$ and $k(u,v)=a<1$. The estimator equals $a-1<0$, whereas the exact empirical MMD is zero because the empirical measures agree. This estimator can be useful for inference, although it is neither a metric nor a pseudometric. Before using a software output as a distance, check which of these quantities it computes.

\section{Related constructions and Sinkhorn divergences}\label{sec:related}
The ten exact distances are metrics, while similar-looking formulas may have different properties. Table~\ref{tab:variants} collects these alternatives and distinguishes failure of separation from failure of the triangle inequality.

\begin{table}[!htbp]
\centering
\small
\setlength{\tabcolsep}{5pt}
\renewcommand{\arraystretch}{1.25}
\begin{tabularx}{\textwidth}{@{}>{\raggedright\arraybackslash}p{4.35cm}>{\centering\arraybackslash}p{1.15cm}Y@{}}
\toprule
\textbf{Construction} & \textbf{Status} & \textbf{Key distinction}\\
\midrule
Finite-direction sliced distance & $\circ$ & Zero may occur for different measures with the same selected projections.\\
\addlinespace[5pt]
Noncharacteristic MMD, finite mean-feature distance & $\circ$ & Different measures can have the same representation.\\
\addlinespace[5pt]
Squared distances, energy or JS without a square root & $\times$ & The triangle inequality fails for the examples discussed here.\\
\addlinespace[5pt]
Regularized transport $T_\varepsilon$ & $\times$ & A measure can have positive cost to itself.\\
\addlinespace[5pt]
Sinkhorn $S_\varepsilon$ and its square root & $\times$ & Removing self-cost does not guarantee the triangle inequality.\\
\addlinespace[5pt]
Cosine-type scores & $\sim$ & Equal inputs have score one, rather than distance zero.\\
\bottomrule
\end{tabularx}
\caption{Related constructions and the metric properties that can fail. The status symbols denote a pseudometric ($\circ$), failure of the metric axioms in general ($\times$), and a similarity ($\sim$). Parameters and representations are fixed. A representation-based pseudometric is a metric on a class exactly when the representation distinguishes that class. The squared-distance entry concerns $W_{2,\mathrm g}^2$, $W_{2,\mathrm c}^2$, $\SW^2$, $\MMD^2$, and $H^2$. The unrooted quantities are $\Energy$ and $\JS$. Equations~\eqref{eq:Teps} and~\eqref{eq:Seps} specify the Sinkhorn convention.}\label{tab:variants}
\end{table}

\subsection{A three-point counterexample for squared distances}
Take point masses at angles $0,\pi/3,2\pi/3$ on a great circle. Their squared chordal distances are $1,1,3$ for the two adjacent pairs and the outer pair. Squared chordal, geodesic, and sliced Wasserstein distances all fail the triangle inequality for this triple. The unrooted cosine dissimilarities $1-\Cot$ and $1-\Csw$ fail as well. The mixture counterexamples in Sections~\ref{sec:energy} to~\ref{sec:js} treat squared MMD, energy, squared Hellinger, and JS.

\subsection{Regularized transport and Sinkhorn divergence}
Regularized transport adds a penalty to the transport cost. Here the penalty discourages a coupling from differing too much from independent pairing, with $\varepsilon$ controlling its weight. The regularized cost can be positive when a measure is compared with itself. Sinkhorn divergence subtracts two self-comparison terms to remove this positive self-cost, giving zero for equal measures. To specify the convention, let $c(x,y)=\norm{x-y}^2/4\in[0,1]$ and fix $\varepsilon>0$. We define
\begin{align}
 T_\varepsilon(\mu,\nu)&=
 \inf_{\gamma\in\Pi(\mu,\nu)}\left\{\int c\dd\gamma+
 \varepsilon\KL(\gamma\Vert\mu\otimes\nu)\right\},\label{eq:Teps}\\
 S_\varepsilon(\mu,\nu)&=T_\varepsilon(\mu,\nu)
 -\tfrac12T_\varepsilon(\mu,\mu)-\tfrac12T_\varepsilon(\nu,\nu).\label{eq:Seps}
\end{align}
Here $\KL(\gamma\Vert\mu\otimes\nu)=\int\log(\mathrm d\gamma/\mathrm d(\mu\otimes\nu))\,d\gamma$ when $\gamma$ has a density relative to $\mu\otimes\nu$, and the penalty is infinite otherwise. The regularized cost satisfies $0\leq T_\varepsilon\leq1$.

Nonnegativity after subtracting the self-comparison terms follows from an external theorem. Theorem~1 of \citet{feydy_2019_InterpolatingOptimalTransport} applies to this cost.

The divergence is nonnegative and vanishes exactly for equal measures. Its bounds are
\begin{equation*}
 0\leq S_\varepsilon\leq1.
\end{equation*}
The regularized self-cost is positive unless the measure is a point mass.

For point masses, $S_\varepsilon(\delta_x,\delta_y)=c(x,y)$, so the three-point example above violates its triangle inequality for every $\varepsilon>0$. Taking a square root does not always restore the triangle inequality. \citet{lavenant_2025_RiemannianGeometrySinkhorn} give counterexamples in Euclidean settings, and the next example establishes a failure for measures supported on the sphere.

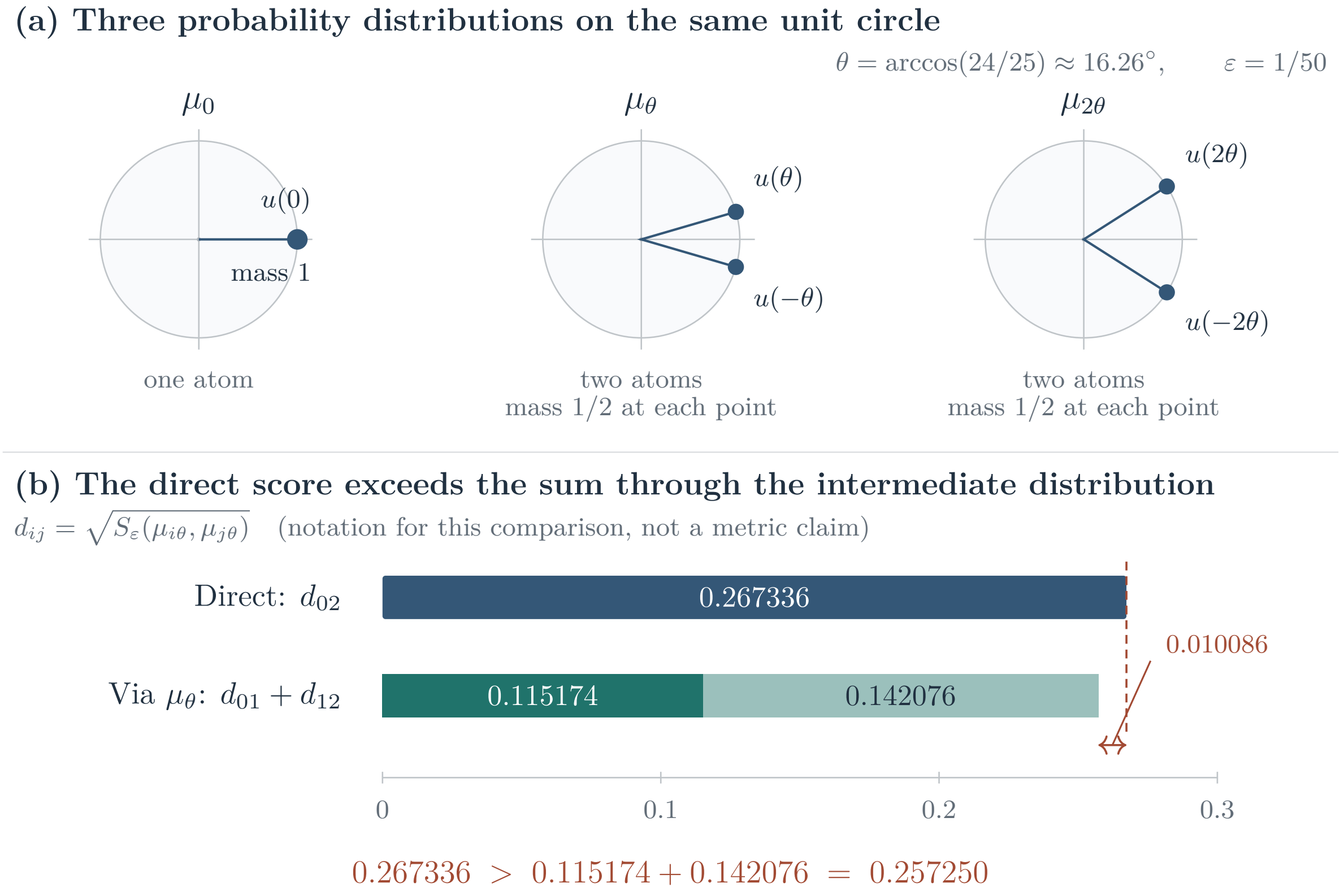
\begin{figure}[!htbp]
\centering
\resizebox{\textwidth}{!}{
\begingroup
\definecolor{sinkink}{HTML}{203040}
\definecolor{sinkmuted}{HTML}{606B76}
\definecolor{sinkblue}{HTML}{345777}
\definecolor{sinkteal}{HTML}{20736B}
\definecolor{sinklight}{HTML}{E8EEF3}
\definecolor{sinkalert}{HTML}{A24A33}
\begin{tikzpicture}[
  x=1cm,y=1cm,text=sinkink,
  font=\rmfamily\fontsize{10.5}{13}\selectfont,
  title/.style={font=\rmfamily\bfseries\fontsize{11}{13}\selectfont},
  small/.style={font=\rmfamily\fontsize{9}{11}\selectfont,text=sinkmuted},
  atom/.style={circle,fill=sinkblue,inner sep=0pt,minimum size=5.5pt},
  guide/.style={draw=sinkmuted!40,line width=.5pt},
]
\path[use as bounding box] (0,-1.2) rectangle (16,9.9);
\node[title,anchor=west] at (0,9.55) {(a) Three probability distributions on the same unit circle};
\node[small,anchor=east] at (16,9.05)
  {$\theta=\arccos(24/25)\approx16.26^\circ,\qquad \varepsilon=1/50$};

\foreach \cx/\name/\subtitle in {2.35/{\mu_0}/{\text{one atom}},7.65/{\mu_\theta}/{\text{two atoms}},12.95/{\mu_{2\theta}}/{\text{two atoms}}}{
  \begin{scope}[shift={(\cx,6.95)}]
    \draw[guide,fill=sinklight!28] (0,0) circle[radius=1.18];
    \draw[guide] (-1.32,0)--(1.36,0);
    \draw[guide] (0,-1.32)--(0,1.32);
    \fill[sinkmuted!50] (0,0) circle[radius=.025];
    \node[font=\rmfamily\bfseries\fontsize{12}{14}\selectfont] at (0,1.62) {$\name$};
    \node[small] at (0,-1.68) {$\subtitle$};
  \end{scope}
}
\begin{scope}[shift={(2.35,6.95)}]
  \draw[draw=sinkblue,line width=.8pt] (0,0)--(1.18,0);
  \node[atom,minimum size=7pt] at (1.18,0) {};
  \node[small,anchor=south east,text=sinkink] at (1.48,.17) {$u(0)$};
  \node[small,anchor=north east,text=sinkink] at (1.48,-.16) {mass $1$};
\end{scope}
\begin{scope}[shift={(7.65,6.95)}]
  \draw[draw=sinkblue,line width=.8pt] (1.1328,.3304)--(0,0)--(1.1328,-.3304);
  \node[atom] at (1.1328,.3304) {};
  \node[atom] at (1.1328,-.3304) {};
  \node[small,anchor=south west,text=sinkink] at (1.21,.43) {$u(\theta)$};
  \node[small,anchor=north west,text=sinkink] at (1.21,-.43) {$u(-\theta)$};
  \node[small] at (0,-2.03) {mass $1/2$ at each point};
\end{scope}
\begin{scope}[shift={(12.95,6.95)}]
  \draw[draw=sinkblue,line width=.8pt] (.994976,.634368)--(0,0)--(.994976,-.634368);
  \node[atom] at (.994976,.634368) {};
  \node[atom] at (.994976,-.634368) {};
  \node[small,anchor=south west,text=sinkink] at (1.08,.71) {$u(2\theta)$};
  \node[small,anchor=north west,text=sinkink] at (1.08,-.71) {$u(-2\theta)$};
  \node[small] at (0,-2.03) {mass $1/2$ at each point};
\end{scope}
\draw[sinkmuted!25] (0,4.40)--(16,4.40);
\node[title,anchor=west] at (0,3.99) {(b) The direct score exceeds the sum through the intermediate distribution};
\node[small,anchor=west] at (0,3.50) {$d_{ij}=\sqrt{S_\varepsilon(\mu_{i\theta},\mu_{j\theta})}$\quad (notation for this comparison, not a metric claim)};

\def\xbase{4.55}
\def\factor{33.3333333333}
\pgfmathsetmacro{\direct}{\xbase+\factor*0.267335993634}
\pgfmathsetmacro{\first}{\xbase+\factor*0.115173766333}
\pgfmathsetmacro{\via}{\xbase+\factor*0.257250164709}
\node[anchor=east] at (4.20,2.66) {Direct: $d_{02}$};
\node[anchor=east] at (4.20,1.48) {Via $\mu_\theta$: $d_{01}+d_{12}$};
\fill[sinkblue,rounded corners=1pt] (\xbase,2.40) rectangle (\direct,2.92);
\node[text=white,font=\rmfamily\bfseries\fontsize{10}{12}\selectfont] at ({(\xbase+\direct)/2},2.66) {$0.267336$};
\fill[sinkteal] (\xbase,1.22) rectangle (\first,1.74);
\fill[sinkteal!45!white] (\first,1.22) rectangle (\via,1.74);
\node[text=white,font=\rmfamily\bfseries\fontsize{10}{12}\selectfont] at ({(\xbase+\first)/2},1.48) {$0.115174$};
\node[text=sinkink,font=\rmfamily\bfseries\fontsize{10}{12}\selectfont] at ({(\first+\via)/2},1.48) {$0.142076$};
\draw[sinkalert,densely dashed,line width=.7pt] (\direct,1.05)--(\direct,3.10);
\draw[sinkalert,line width=.7pt,<->] (\via,.89)--(\direct,.89);
\node[small,anchor=west,text=sinkalert] at (13.80,2.10) {$0.010086$};
\draw[sinkalert,line width=.6pt] (13.75,1.90)--({(\via+\direct)/2},.89);
\draw[guide] (\xbase,.50)--(14.55,.50);
\foreach \v in {0,0.1,0.2,0.3}{
  \pgfmathsetmacro{\xx}{\xbase+\factor*\v}
  \draw[guide] (\xx,.43)--(\xx,.57);
  \node[small,anchor=north] at (\xx,.35) {$\v$};
}
\node[font=\rmfamily\bfseries\fontsize{11}{13}\selectfont,text=sinkalert] at (8,-.65)
 {$0.267336\;>\;0.115174+0.142076\;=\;0.257250$};
\end{tikzpicture}
\endgroup}
\caption[The square-root Sinkhorn counterexample]{The counterexample in Example~\ref{ex:sinkhorn}, with $\theta=\arccos(24/25)$ and $\varepsilon=1/50$. \textbf{(a)} The three distributions on the unit circle. At angle zero the atoms coincide, giving one atom of mass one. Each other distribution has two atoms of mass $1/2$. \textbf{(b)} The direct square-root Sinkhorn score exceeds the sum through the intermediate distribution. Bar lengths represent numerical scores rather than paths along the circle. Rounded values illustrate a violation established by exact inequalities.}\label{fig:sinkhorn-counterexample}
\end{figure}

\begin{example}[Square-root Sinkhorn fails on the circle]\label{ex:sinkhorn}
Let $u(s)=(\cos s,\sin s)$ and $\mu_s=(\delta_{u(s)}+\delta_{u(-s)})/2$. Set $\theta=\arccos(24/25)$ and $\varepsilon=1/50$. For the three distributions with parameters $0,\theta,2\theta$, write $S_{ij}=S_\varepsilon(\mu_{i\theta},\mu_{j\theta})$. Then
\[
 \sqrt{S_{02}}>\sqrt{S_{01}}+\sqrt{S_{12}},
\]
so square-root Sinkhorn fails the triangle inequality. Exact formulas and rational bounds establish this strict inequality. Adding zero coordinates gives the same example in every $d\geq2$. Figure~\ref{fig:sinkhorn-counterexample} shows the distributions and compares the direct score with the sum through the intermediate distribution.
\end{example}

\FloatBarrier

\section{Conclusion}\label{sec:conclusion}
This article compares ten metrics for probability distributions on a unit sphere, including finite samples of directions and continuous angular distributions. Consistent definitions and a common maximum value of one allow their properties to be compared directly. Their shared range does not make them interchangeable, because they emphasize different aspects of a distribution and respond differently to its changes.

One distinction concerns geometric proximity and exact probability overlap. Full and sliced Wasserstein distances, their associated angles, Gaussian MMD, and chordal energy describe weak convergence. They recognize, for example, that moving a point mass by a small angle leaves it close to its starting point. Total variation, Hellinger, and Jensen-Shannon distance describe the stronger total-variation convergence. They assign distance one to every mutually singular pair, including arbitrarily close distinct point masses. In the scattering example, this distinction determines whether nearby but nonidentical recorded directions count as similar.

A second distinction concerns vector representations. Sliced Wasserstein, Gaussian MMD, chordal energy, Hellinger, and Jensen-Shannon distance admit exact Hilbert norm representations. The angular distances are also metrics, although angles between Hilbert-space vectors need not be Hilbert norm distances. Likewise, optimal cosine matching and positive semidefiniteness impose different requirements on a similarity. The identities and comparison bounds specify which constructions share convergence or rankings and where those agreements end.

The guarantees of an exact distance also need to be checked for its implementation. Finite projections or feature summaries may lose separation, and squaring a distance may remove its triangle inequality. An unbiased estimator can be negative on particular samples. Subtracting regularized transport self-costs does not ensure a metric, even after taking a square root. These constructions can still be useful, provided their remaining mathematical properties match the intended use.

The article is a mathematical guide rather than an empirical ranking of methods. It neither selects a universally best metric nor models detector error or establishes finite-sample accuracy for a particular experiment. Its table, diagrams, proofs, and counterexamples provide a basis for those later choices. A comparison should preserve the information relevant to the problem, and the construction should have the required properties. Those properties must then be checked for the numerical approximation actually used.

\bibliographystyle{abbrvnat}
\bibliography{Bibliography}

\clearpage
\appendix
\section{Appendix: Proofs and derivations}\label{app:proofs}

\subsection{Average directions}\label{app:mean-direction}
\begin{proof}[Verification of the example in Section~\ref{sec:table}]
Averaging $u_+$ and $u_-$ cancels their first coordinates and yields $\cos\alpha\,z$. Since $\cos\alpha>0$, normalization gives $z$, the mean direction of $\delta_z$.
\end{proof}

\subsection{Vector representations}\label{app:vector}

\begin{proof}[Proof of Lemma~\ref{lem:cnd}]
For a Hilbert representation and coefficients summing to zero, the left side of~\eqref{eq:cnd} is $-2\norm{\sum_i a_i\Phi(z_i)}^2$. Conversely, fix $o\in\mathcal X$ and set
\[
 K_o(z,z')=\tfrac12\{D(z,o)^2+D(z',o)^2-D(z,z')^2\}.
\]
For any finite collection of coefficients, add the point $o$ with a coefficient equal to minus their sum. The enlarged collection has coefficients summing to zero, so~\eqref{eq:cnd} shows that $K_o$ is positive semidefinite. To construct its Hilbert space, assign a formal vector $e_z$ to each point and set $\langle e_z,e_{z'}\rangle=K_o(z,z')$. Extend this rule to finite linear combinations, identify vectors whose difference has zero norm, and add the limits of all Cauchy sequences. This gives a Hilbert space in which $\|e_z-e_{z'}\|^2=D(z,z')^2$. Different points therefore have different representations.

For a Hilbert representation, write the Gaussian kernel as
$e^{-t\norm{u}^2}e^{-t\norm{v}^2}e^{2t\ip uv}$. The last factor is positive semidefinite by its power series and tensor powers, and finite Gram matrices preserve positivity under the limit. Conversely, if all these Gaussian kernels are positive semidefinite, differentiating their quadratic forms at $t=0$ for coefficients with zero sum gives~\eqref{eq:cnd}. Finally, coefficients $(1,-1,1,-1)$ on the four-cycle give $4b^2-8a^2$.
\end{proof}

\begin{proof}[Proof of Lemma~\ref{lem:angle}]
The only nonimmediate property is the triangle inequality. Put $a=\alpha(u,v)$ and $b=\alpha(v,w)$. Split $u$ and $w$ into parts parallel and perpendicular to $v$. Their parallel parts contribute $\cos a\cos b$, and Cauchy-Schwarz bounds the inner product of the perpendicular parts below by $-\sin a\sin b$. Hence
$\ip uw\geq\cos a\cos b-\sin a\sin b=\cos(a+b)$.
If $a+b\leq\pi$, applying the decreasing function $\arccos$ gives the result. Otherwise, the angle is already bounded by $\pi<a+b$. Formula~\eqref{eq:anglechord} follows from $\norm{u-v}^2=2-2\cos\alpha$. Four equally spaced vectors on a two-dimensional great circle have normalized adjacent angle $1/2$ and opposite angle $1$, contradicting Lemma~\ref{lem:cnd}.
\end{proof}

\subsection{Geodesic and chordal Wasserstein metrics}\label{app:wasserstein}

\begin{proof}[Proof of Theorem~\ref{thm:Wp}]
\emph{Existence and zero distance.} Independent draws from $\mu$ and $\nu$ provide at least one coupling. A standard compactness result gives a convergent subsequence from any sequence of couplings on the compact product sphere, and the limit has the same prescribed marginals. Since the cost is bounded and continuous, a plan of minimum cost exists. Nonnegativity follows from the cost, and symmetry follows by reversing a plan. Pairing each point with itself gives zero cost when the measures agree. Conversely, zero cost forces $X=Y$ with probability one, so the two distributions must agree.

\emph{Triangle inequality.} Combine optimal plans for $(\mu,\nu)$ and $(\nu,\lambda)$ into a joint distribution of $(X,Y,Z)$. One construction first draws $Y$ from $\nu$, then draws $X$ and $Z$ independently conditional on $Y$, using the conditional distributions from the two plans. This preserves both required pairwise plans and is called \emph{gluing}. The triangle inequality for $\rho$ and the triangle inequality for the $L^p$ norm (Minkowski's inequality) give
\[
 W_{p,\rho}(\mu,\lambda)
 \leq\norm{\rho(X,Z)}_{L^p}
 \leq\norm{\rho(X,Y)}_{L^p}+\norm{\rho(Y,Z)}_{L^p}.
\]
This proves the triangle inequality and completes the metric argument.

\emph{Maximum distance.} No pair of sphere points is farther apart than $B$, so every coupling has cost at most $B^p$. Opposite point masses reach this value. If the minimum cost equals $B^p$, even independent draws must be opposite with probability one. Thus, for $\mu$-almost every $x$, we have $\nu(\{-x\})=1$. Fixing one such $x$ shows that both measures must be point masses at opposite points.

\emph{Convergence.} First suppose $W_{p,\rho}(\mu_n,\mu)\to0$ and choose optimal couplings. We show that averages of every continuous function $h$ converge. Given $\epsilon>0$, uniform continuity gives $\delta>0$ such that $|h(x)-h(y)|<\epsilon$ whenever $\rho(x,y)<\delta$. Split the coupled pairs into those less than $\delta$ apart and the remaining pairs. Markov's inequality bounds the probability of the second group, giving
\[
 \left|\int h\dd\mu_n-\int h\dd\mu\right|
 \leq\epsilon+2\norm h_\infty\delta^{-p}W_{p,\rho}(\mu_n,\mu)^p.
\]
Letting $n$ tend to infinity and then $\epsilon$ tend to zero proves weak convergence.

Conversely, suppose $\mu_n\Rightarrow\mu$. Divide the sphere into finitely many measurable pieces, each of diameter at most $\delta$, whose boundaries have zero $\mu$-mass. To construct such a partition, cover the sphere with finitely many small open balls whose boundaries have zero mass, then remove overlaps in order. Weak convergence implies that the probability of each piece converges. Within each piece, match as much mass as possible between the two measures, so those pairs move at most $\delta$. Match the remaining mass in any way. Its amount $r_n$ tends to zero, so
$W_{p,\rho}^p\leq\delta^p+B^p r_n$.
Take $n\to\infty$ and then $\delta\downarrow0$.
\end{proof}

\begin{proof}[Ground geometries and transport orders]
For $\theta=\rho_{\mathrm g}(x,y)\in[0,\pi]$,
\[
 \rho_{\mathrm c}(x,y)=2\sin(\theta/2),\qquad
 \frac2\pi\theta\leq2\sin(\theta/2)\leq\theta.
\]
Applying these bounds inside every coupling gives the ground-geometry comparisons in Section~\ref{sec:wasserstein}.

For a fixed ground metric and $1\leq p<q<\infty$, write $D_p=W_{p,\rho}/B$. For the first inequality, apply the ordering of $L^p$ and $L^q$ norms under a probability distribution to an optimal $q$-coupling. For the other inequality, use $\rho^q\leq B^{q-p}\rho^p$ in an optimal $p$-coupling.
\end{proof}

\begin{proof}[Non-Hilbertian Wasserstein families]
For geodesic Wasserstein, restrict to four point masses at successive quarter-turns of a great circle. Every $p$ gives normalized adjacent distances of $1/2$ and opposite distances of $1$, so Lemma~\ref{lem:cnd} rules out a Hilbert representation.

For chordal Wasserstein, use the antipodal measures defined in Section~\ref{sec:wasserstein}.
There are two ways to match one antipodal pair to the other point by point. In each matching, both paired points have the same distance. Every allowed two-by-two coupling is a weighted average of these two matchings, so one of them minimizes the cost. For every finite $p\geq1$, the minimum cost gives~\eqref{eq:antipodalWp}.
For normalized chordal Wasserstein, adjacent squared distances are $(2-\sqrt2)/4$ and opposite squared distances are $1/2$. Use the coefficients $(1,-1,1,-1)$ in~\eqref{eq:cnd}. The resulting sum, $2(\sqrt2-1)>0$, violates the Hilbertian criterion. Thus chordal Wasserstein is not Hilbertian on the full space of measures for any stated value of $p$.
\end{proof}

\subsection{Optimal cosine matching and the OT angle}\label{app:otangle}

\begin{proof}[The cosine identity]
The identity $\norm{x-y}^2=2-2x^\top y$ converts the maximization in~\eqref{eq:cot} into squared chordal transport minimization and gives~\eqref{eq:cotidentity}.
\end{proof}

\begin{proof}[Proof of Theorem~\ref{thm:otangle}]
Use gluing to combine couplings maximizing~\eqref{eq:cot} for $(\mu,\nu)$ and $(\nu,\lambda)$. This gives random vectors $(X,Y,Z)$ on a common probability space. Regard them as vectors in $L^2(\Omega;\R^d)$, where the inner product of $X$ and $Y$ is $\E[X^\top Y]$. Each has norm one because $\E\|X\|^2=1$, and likewise for $Y,Z$. Their angles satisfy Lemma~\ref{lem:angle}. The resulting $(X,Z)$ pair is one allowed coupling, so $\Cot(\mu,\lambda)\geq\E[X^\top Z]$. Since $\arccos$ is decreasing, this proves the triangle inequality. Equation~\eqref{eq:cotidentity} shows that zero angle occurs only for equal measures, and symmetry and the range bounds follow directly.

The transformation in~\eqref{eq:aot} is continuous, strictly increasing, and zero only at zero. Therefore it preserves convergence to zero, and Theorem~\ref{thm:Wp} gives the weak topology and the diameter assertion. Its restriction to four quarter-turn point masses is the non-Hilbertian angular four-cycle. The first pair of bounds in~\eqref{eq:aotbounds} follows from $t\leq\arcsin t\leq\pi t/2$ on $[0,1]$.

For the final bound, consider $F(s)=\cos\sqrt s$ on $[0,\pi^2]$. Its convexity follows from the second derivative, evaluated at $s=\theta^2>0$,
\[
 F''(s)=\frac{\sin\theta-\theta\cos\theta}{4\theta^3}\geq0,
\]
because the numerator starts at zero and has derivative $\theta\sin\theta\geq0$. In a coupling minimizing squared geodesic cost, Jensen's inequality gives
$\Cot\geq\E\cos\rho_{\mathrm g}(X,Y)\geq\cos W_{2,\mathrm g}$.
Apply $\arccos$ and divide by $\pi$.
\end{proof}

\begin{proof}[Indefiniteness of the OT cosine]
The cosine $\Cot$ does not always give a positive semidefinite similarity matrix. For the antipodal measures in~\eqref{eq:antipodalWp}, its entries are $|u_i^\top u_j|$. The matrix has eigenvector $(1,-1,1,-1)$ with negative eigenvalue $1-\sqrt2$. Rescaling the scores to $(1+\Cot)/2$ gives the negative eigenvalue $(1-\sqrt2)/2$, because the added all-ones matrix sends that vector to zero.
\end{proof}

\subsection{Sliced Wasserstein and its angle}\label{app:sliced}

\begin{proof}[Measurability and optimality of quantile matching]
The quantiles are measurable jointly in $(\theta,t)$, because the projected distribution functions are measurable and rational thresholds describe the sets where a quantile is below a given value.

The optimality of quantile matching follows from a simple bound on joint tail probabilities. With the two marginal distributions fixed, minimizing $\E(U-V)^2$ is equivalent to maximizing $\E[(U+1)(V+1)]$. Since $U+1,V+1\geq0$, this expectation equals the integral of $\Pr(U+1>s,V+1>t)$ over nonnegative $s,t$. Each joint probability is at most the smaller of the two marginal probabilities. Drawing one uniform variable $T\in(0,1)$ and setting $U=Q_\mu(T)$, $V=Q_\nu(T)$ reaches this upper bound for every $s,t$. This proves the formula, including for distributions with point masses.
\end{proof}

\begin{proof}[Proof of Theorem~\ref{thm:sw}]
Integrating the square of a quantile over $t\in(0,1)$ gives the second moment of the corresponding distribution. Uniform directions satisfy $\int\theta\theta^\top\dd\sigma=I_d/d$ because all coordinate directions are treated equally and the trace is one. Therefore
\[
 \norm{\Phi_\mu}^2
 =d\int\int (\theta^\top x)^2\dd\mu(x)\dd\sigma(\theta)
 =\int\norm{x}^2\dd\mu(x)=1.
\]
Equations~\eqref{eq:swembed}, \eqref{eq:csw}, and~\eqref{eq:asw} follow by expanding a squared Hilbert norm and using Lemma~\ref{lem:angle}.

To prove injectivity, zero sliced distance implies $\mu_\theta=\nu_\theta$ for almost every direction. Coupling two projections of the same $X$ gives
\[
 W_2(\mu_\theta,\mu_\eta)\leq\norm{\theta-\eta}.
\]
The triangle inequality therefore makes $\theta\mapsto W_2(\mu_\theta,\nu_\theta)$ continuous. Every nonempty open patch of directions has positive $\sigma$-probability. A continuous function that is zero almost everywhere must consequently be zero in every direction. All projected distributions therefore agree, which determines the original measure by the Cram\'er-Wold principle \citep{cramer_1936_TheoremsDistributionFunctions}. To see the argument directly, use the characteristic function $\widehat\mu(\omega)=\int e^{\mathrm i\omega^\top x}\,d\mu(x)$, the Fourier transform of a probability measure. Its values at $\omega=t\theta$ are determined by the projection in direction $\theta$. Thus the characteristic functions agree everywhere, and the uniqueness theorem for characteristic functions gives $\mu=\nu$.

For weak convergence, $\mu_n\Rightarrow\mu$ implies weak convergence of every projected measure. The compact-interval argument of Theorem~\ref{thm:Wp} gives projected $W_2\to0$, and dominated convergence applies because projected squared distances are at most four. Conversely, every subsequence of $\mu_n$ has a further weakly convergent subsequence by compactness of $\Prob$ in the weak topology. If $\Dsw(\mu_n,\mu)\to0$, continuity and separation force every such subsequential limit to be $\mu$. The entire sequence therefore converges weakly. Equation~\eqref{eq:asw} gives the same conclusion for the angle.

For point masses, $Q_{\delta_x}(\theta,t)=\theta^\top x$, and thus
\begin{equation}\label{eq:swdirac}
 \Csw(\delta_x,\delta_y)=x^\top y,
 \quad\Dsw(\delta_x,\delta_y)=\tfrac12\norm{x-y},
 \quad\Asw(\delta_x,\delta_y)=\rho_{\mathrm g}(x,y)/\pi.
\end{equation}
Antipodal point masses establish the sharp diameter bound. Equation~\eqref{eq:swembed} proves the Hilbertian claim, and~\eqref{eq:csw} gives a Gram kernel. Equation~\eqref{eq:swdirac} and the angular four-cycle rule out a Hilbert representation of $\Asw$.
\end{proof}

\begin{proof}[Proof of Proposition~\ref{prop:swbounds}]
The difference between projected means is $\theta^\top(m_\mu-m_\nu)$, whose squared value is no larger than projected $W_2^2$ by Jensen's inequality in a coupling. Integration over uniform directions gives the claimed lower bound. For any full coupling $\gamma$, its projected coupling is feasible, so
\[
 \SW^2\leq\iint\int(\theta^\top(x-y))^2\dd\sigma(\theta)\dd\gamma(x,y)
 =\frac1d\int\norm{x-y}^2\dd\gamma.
\]
Minimize over $\gamma$ and use the monotone chord-to-angle relation and the cosine identities.
\end{proof}

\begin{proof}[Perpendicular antipodal measures and maximum sliced distance]
For the perpendicular antipodal measures in Section~\ref{sec:sliced}, every cross inner product is zero, so $\Cot=0$. Their projected quantiles have magnitudes $|\cos\theta|$ and $|\sin\theta|$, giving
\[
 \Csw=2\int_0^{2\pi}|\cos\theta\sin\theta|\frac{\mathrm d\theta}{2\pi}=\frac2\pi>0.
\]
Thus full and sliced matching agree for individual point masses but can disagree for distributions. Equations~\eqref{eq:swbounds} and Theorem~\ref{thm:Wp} also show that sliced Wasserstein and its angle reach one only for point masses at opposite points.
\end{proof}

\subsection{Gaussian maximum mean discrepancy}\label{app:gaussian}

\begin{proof}[Proof of Theorem~\ref{thm:mmd}]
We first check that the kernel gives valid inner products and distinguishes measures, using the characteristic function of a Gaussian distribution.
\[
 k_\tau(x,y)=\int_{\R^d}e^{\mathrm i\omega^\top(x-y)}\dd\Lambda_\tau(\omega),
 \qquad\Lambda_\tau=N(0,\tau I_d).
\]
For arbitrary finite coefficients, the resulting quadratic form is an integral of a squared absolute value, so the kernel is positive semidefinite. Applying this representation to $\eta=\mu-\nu$ gives
\begin{equation}\label{eq:gaussfourier}
 \MMD_\tau^2(\mu,\nu)=\int_{\R^d}|\widehat\mu(\omega)-\widehat\nu(\omega)|^2\dd\Lambda_\tau(\omega).
\end{equation}
The Gaussian distribution $\Lambda_\tau$ has positive density everywhere. If the integral is zero, continuity forces the two characteristic functions to agree everywhere, so $\mu=\nu$. The same reasoning works for finite signed measures, which allow negative as well as positive mass. This stronger injectivity result justifies the normalized kernel-mean construction in Section~\ref{sec:gaussian}.

Since $e^{-2\tau}\leq k_\tau\leq1$, formula~\eqref{eq:mmddef} gives
$0\leq\MMD_\tau^2\leq2(1-e^{-2\tau})$.
Antipodal point masses attain the upper bound. Conversely, equality requires both self expectations to be one. Since $k_\tau(x,y)=1$ only at $x=y$, each measure must then be a point mass, and the cross expectation forces antipodality.

The norm representation proves the remaining metric axioms and Hilbertian structure. If $\mu_n\Rightarrow\mu$, continuity of the bounded kernel and weak convergence of product measures imply $\MMD_\tau(\mu_n,\mu)\to0$. The converse follows from weak compactness and separation, using the subsequence argument in Theorem~\ref{thm:sw}.
\end{proof}

\begin{proof}[Bounds through transport and total variation]
The feature distance obeys
\[
 \norm{\phi_\tau(x)-\phi_\tau(y)}^2
 =2(1-e^{-\tau\norm{x-y}^2/2})\leq\tau\norm{x-y}^2.
\]
Under any coupling, the norm of the average feature difference is no larger than the average feature distance. Minimizing the resulting bound on average chordal displacement gives the normalized transport bound in Section~\ref{sec:gaussian}.
Linearity of the mean embedding gives the further bound $D_\tau\leq\TV$ by the common-mass comparison~\eqref{eq:tvkernel}.
\end{proof}

\begin{proof}[Limit as the Gaussian parameter vanishes]
The expansion $k_\tau(x,y)=1-\tau(1-x^\top y)+O(\tau^2)$ holds uniformly for sphere points. Substituting it into~\eqref{eq:mmddef} gives the first identity in~\eqref{eq:mmdsmalltau}. The normalization satisfies $2(1-e^{-2\tau})=4\tau+O(\tau^2)$, giving the stated limit.
\end{proof}

\begin{proof}[Normalized kernel means]
The positive Gaussian kernel makes the denominator of $C_\tau$ nonzero. Its numerator is at least $e^{-2\tau}$, and each squared mean norm is at most one. Cauchy-Schwarz gives the upper bound on the normalized cosine. Indeed, if $m_k(\mu)=c\,m_k(\nu)$ for $c>0$, applying~\eqref{eq:gaussfourier} to the signed measure $\mu-c\nu$ gives $\mu=c\nu$. Since both measures have total mass one, $c=1$, which proves injectivity of the normalized mean map. The chord and angle metrics then follow from the unit-vector representations.
\end{proof}

\subsection{Chordal energy}\label{app:energy}

\begin{proof}[Proof of Theorem~\ref{thm:energy}]
We express ordinary Euclidean distance as an integral of Gaussian similarity differences. For $r\geq0$, integration by parts and the Gaussian integral give
\begin{equation}\label{eq:laplacenorm}
 r=\frac1{\sqrt{2\pi}}\int_0^\infty(1-e^{-\tau r^2/2})\tau^{-3/2}\dd\tau.
\end{equation}
After setting $a=r^2/2$, the integral is $2\sqrt\pi\sqrt a$. Let $\eta=\mu-\nu$, so that $\eta(\Sph)=0$ and $\Energy=-\iint\norm{x-y}\dd\eta(x)\dd\eta(y)$. Substituting~\eqref{eq:laplacenorm} cancels the constant term and leaves the Gaussian term $\MMD_\tau^2$, proving~\eqref{eq:energymixture}. Absolute integrability justifies the interchange of integrals. On the sphere, $1-e^{-\tau r^2/2}\leq\min(2\tau,1)$, whose product with $\tau^{-3/2}$ is integrable.

The same identity proves that $\sum_{i,j}a_i a_j\|x_i-x_j\|\leq0$ whenever $\sum_i a_i=0$. This sum is minus an integral of nonnegative Gaussian-kernel quadratic forms, which establishes conditional negative definiteness of Euclidean distance itself. \citet{lyons_2013_DistanceCovarianceMetric} develops the role of negative-type distances in comparisons of distributions. Add the origin to the points $x_1,\ldots,x_n$, giving it a coefficient that makes the total sum zero. Rearranging the inequality shows that
$\tfrac12(\norm{x}+\norm{y}-\norm{x-y})$
is a positive semidefinite kernel, equal to $k_{\mathrm E}$ on the unit sphere. Expanding the MMD of this kernel gives~\eqref{eq:energykernel}. If $\mu\ne\nu$, every Gaussian MMD in~\eqref{eq:energymixture} is positive, so the integral is positive as well. The energy kernel therefore distinguishes measures, and $\DE=\MMD_{k_{\mathrm E}}/\sqrt2$ is a Hilbertian metric.

Equation~\eqref{eq:energydef} gives $\Energy\leq4$, with equality for antipodal point masses. Equality requires zero self distances and maximal cross distance, so only antipodal point masses attain the diameter. Since $k_{\mathrm E}$ is bounded and continuous, the compactness argument used for Gaussian MMD proves weak metrization.

Its feature vectors satisfy the exact identity $\norm{\phi_{\mathrm E}(x)-\phi_{\mathrm E}(y)}^2=\norm{x-y}$. In any coupling,
\[
 \MMD_{k_{\mathrm E}}\leq\int\sqrt{\norm{x-y}}\dd\gamma
 \leq\left(\int\norm{x-y}\dd\gamma\right)^{1/2}.
\]
Minimizing gives $\Energy\leq2W_{1,\mathrm c}$, proving the first bound in~\eqref{eq:energybounds}. The second follows from the common-mass comparison~\eqref{eq:tvkernel}, which uses linearity of the mean representation.
\end{proof}

\begin{proof}[Geodesic energy and unrooted energy]
Using geodesic distance in the energy formula can also give zero to different measures. For the two perpendicular antipodal distributions in Section~\ref{sec:sliced}, every angle between the distributions is $\pi/2$. The expected angle between two independent draws from either one is also $\pi/2$, so their geodesic energy is zero. Geodesic Wasserstein has a separate metric proof based on optimal matching. Independent averages and optimal matching are different constructions, and each needs its own proof of separation.

Finally, the unrooted $\Energy$ is not generally a metric. If $\lambda=(\mu+\nu)/2$, linearity of kernel means gives
$\Energy(\mu,\lambda)=\Energy(\lambda,\nu)=\Energy(\mu,\nu)/4$.
For $\mu\ne\nu$, their sum is smaller than the direct energy. The same observation applies to squared MMD.
\end{proof}

\subsection{Total variation and common mass}\label{app:tv}

\begin{proof}[Reference invariance and event probabilities]
Using the sum of two reference measures as a common reference makes the density-change factors cancel in the integral.

The supremum over events is attained by the set where $f\geq g$. The positive and negative parts of $f-g$ have equal integrals because the two densities each integrate to one.
\end{proof}

\begin{proof}[Proof of Proposition~\ref{prop:tv}]
For three measures, use their sum as the common reference measure. Then TV is half the $L^1$ distance between densities. The $L^1$ triangle inequality proves the required triangle inequality, and zero distance forces equal densities. Since
\[
 V=1-\int\min(f,g)\dd r,
\]
its range is $[0,1]$, and equality at one is equivalent to no common mass, or mutual singularity. Distinct point masses have distance one, even when their locations converge, so weak convergence does not imply TV convergence.

To rule out a Hilbert representation, choose four distinct sphere points and the following probability vectors on them.
\[
 (\tfrac12,0,\tfrac12,0),\quad
 (0,\tfrac12,\tfrac12,0),\quad
 (0,\tfrac12,0,\tfrac12),\quad
 (\tfrac12,0,0,\tfrac12).
\]
Adjacent TV distances are $1/2$ and opposite distances are one. Lemma~\ref{lem:cnd} rules out a Hilbert embedding.
\end{proof}

\begin{proof}[Common-mass comparisons]
Several bounds in the table follow by separating the probability mass that the two measures share from their residual mass. Put $t=V(\mu,\nu)$ and let $\alpha$ have density $\min(f,g)$ relative to $r$. This common part has mass $1-t$, and when $t>0$, dividing each residual part by $t$ gives probability measures
\[
 \mu'=\frac{\mu-\alpha}{t},\qquad \nu'=\frac{\nu-\alpha}{t}.
\]
Now $\mu-\nu=t(\mu'-\nu')$. In a transport plan, leave the common mass in place and move only the remaining mass $t$. If no two points are farther apart than $B$, the residual cost is at most $tB^p$, proving~\eqref{eq:tvtransport}.
For a norm distance based on a \emph{linear} representation of measures, the difference between representations is also multiplied by $t$. If the maximum distance is one, this gives
$D(\mu,\nu)=tD(\mu',\nu')\leq t$.
Applying this inequality to the Gaussian and energy mean maps gives~\eqref{eq:tvkernel}.
The argument requires linearity, so it does not apply to the nonlinear sliced quantile, square-root density, and Jensen-Shannon representations. Equations~\eqref{eq:tvtransport} and~\eqref{eq:tvkernel} link full transport, Gaussian MMD, and energy distance to total variation.
\end{proof}

\subsection{Hellinger distance}\label{app:hellinger}

\begin{proof}[Proof of Theorem~\ref{thm:hellinger}]
As for TV, changing the common reference measure leaves the answer unchanged because the density-change factors cancel.

For any triple, a common dominating measure represents $H$ as $1/\sqrt2$ times an $L^2$ distance between square-root densities. This proves separation and the triangle inequality. Formula~\eqref{eq:hellinger} and Cauchy-Schwarz give $0\leq H\leq1$, with equality at one precisely when $fg=0$ almost everywhere.

A Hilbert representation for \emph{all} probability measures needs one further justification because a finite reference measure need not dominate the whole space $\Prob$. Each finite collection, however, is dominated by its sum. The square-root densities of that collection give $L^2$ distances, so they satisfy~\eqref{eq:cnd}. Since the condition holds for every finite collection, Lemma~\ref{lem:cnd} gives a Hilbert representation of the entire space without requiring surface densities.

Pointwise, $(\sqrt f-\sqrt g)^2\leq|f-g|$, proving the first comparison. Factoring $|f-g|=|\sqrt f-\sqrt g|(\sqrt f+\sqrt g)$ and applying Cauchy-Schwarz gives
\[
 V\leq\frac12\left(2H^2\right)^{1/2}
 \left(4-2H^2\right)^{1/2}=H\sqrt{2-H^2}.
\]
These inequalities give the topology assertion. As for TV, distinct converging point masses have constant distance one.
\end{proof}

\begin{proof}[Squared Hellinger counterexample]
For the pair of distinct point masses, the affinity is zero. Each point mass has affinity $1/\sqrt2$ with their equal mixture. Substituting these values into~\eqref{eq:hellinger} gives the counterexample in Section~\ref{sec:hellinger}.
\end{proof}

\begin{samepage}
\subsection{Jensen-Shannon distance}\label{app:js}

\begin{lemma}[A scalar squared-distance identity]\label{lem:jsintegral}
For $a,b\geq0$, put
\[
 j(a,b)=\tfrac12\left[a\log\frac{2a}{a+b}+b\log\frac{2b}{a+b}\right],\qquad j(0,0)=0.
\]
Then
\begin{equation*}
 j(a,b)=\frac14\int_0^\infty\frac{(e^{-ta}-e^{-tb})^2}{t^2}\dd t.
\end{equation*}
\end{lemma}
\end{samepage}

\begin{proof}[Proof of Lemma~\ref{lem:jsintegral}]
For $a,b>0$, the numerator is $O(t^2)$ at zero and bounded at infinity. Integration by parts gives the integral without the factor $1/4$ as
\[
 2a\int_0^\infty\frac{e^{-(a+b)t}-e^{-2at}}t\dd t
 +2b\int_0^\infty\frac{e^{-(a+b)t}-e^{-2bt}}t\dd t.
\]
The identity $\int_0^\infty(e^{-ut}-e^{-vt})\,\mathrm dt/t=\log(v/u)$ follows by writing the integrand as $\int_u^v e^{-st}\dd s$ and interchanging the integrals. Substitution gives $4j(a,b)$, and zero arguments follow by a limit. For bounded ranges of $a,b$, a constant bounds the integrand near zero and $t^{-2}$ bounds it at infinity, justifying dominated convergence.
\end{proof}

\begin{proof}[Proof of Theorem~\ref{thm:js}]
Both measures have densities relative to $m$, bounded above by two, so JS is finite even when the original measures have disjoint supports.

For densities relative to $r$, $\JS=\int j(f,g)\dd r$. By Lemma~\ref{lem:jsintegral}, define
\[
 \Psi_f(x,t)=\frac{e^{-tf(x)}-1}{2t}
 \quad\text{in }L^2(r\otimes\mathrm{Leb}_{(0,\infty)}).
\]
Its squared norm is $\int j(f,0)\dd r=(\log2)/2$, and
$\norm{\Psi_f-\Psi_g}^2=\JS(\mu,\nu)$.
For three measures, use their sum as $r$, so the norm triangle inequality proves the triangle inequality for $\sqrt{\JS}$. Equal exponential representations force equal densities, giving separation. For any finite collection, the same representation verifies~\eqref{eq:cnd}. Lemma~\ref{lem:cnd} therefore gives a Hilbert representation for all measures, as for Hellinger distance. Changing the reference measure leaves $\int j(f,g)\,dr$ unchanged because $j(ca,cb)=c\,j(a,b)$ cancels the density-change factors.

To prove the bounds directly, set $s=f+g$ and $u=(f-g)/s$ where $s>0$. With
\[
 h(u)=\tfrac12\{(1+u)\log(1+u)+(1-u)\log(1-u)\},
\]
we have $j(f,g)=s h(u)/2$. The function is even, with $h(0)=h'(0)=0$, $h''(u)=1/(1-u^2)\geq1$, and $h(1)=\log2$. Hence
$u^2/2\leq h(u)\leq(\log2)|u|$.
The upper inequality follows from convexity below the chord on $[0,1]$.
Since $\int s\dd r=2$ and $V=\tfrac12\int s|u|\dd r$, Cauchy-Schwarz gives
\[
 \JS\geq\tfrac14\int s u^2\dd r
 \geq\tfrac14\frac{(\int s|u|\dd r)^2}{\int s\dd r}
 =\tfrac12 V^2,
\]
and the upper bound follows directly. Also $h(u)\leq\log2$, with equality precisely when $|u|=1$, so $\JS=\log2$ precisely when $fg=0$ almost everywhere. This proves the diameter and singularity claims. Equations~\eqref{eq:Jtv} and~\eqref{eq:Htv} establish equality of the three topologies, while distinct converging point masses rule out weak metrization.
\end{proof}

\begin{proof}[Unrooted Jensen-Shannon counterexample]
Substitute the two-point probability vectors into~\eqref{eq:jsdef}. Each point mass has divergence $\tfrac34\log(4/3)$ from their equal mixture, whereas the divergence between the point masses is $\log2$. The two-step sum is therefore strictly smaller.
\end{proof}

\subsection{Comparisons and convergence}\label{app:connections}

\begin{proof}[Common orthogonal transformations]
For full transport, $(x,y)\mapsto(Ux,Uy)$ is a bijection between coupling sets that preserves both ground costs, so the OT angle is also invariant. For slicing, $(U_\#\mu)_\theta=\mu_{U^\top\theta}$, and $\sigma$ is rotation invariant. Gaussian and energy kernels depend only on chordal distance. For TV, Hellinger, and JS, pushing the common dominating measure forward by the same bijection leaves all integrals unchanged. These arguments require a \emph{common} rotation of the two measures. They do not establish invariance under separate alignment.
\end{proof}

\begin{proof}[Antipodal contamination formulas]
Because the first measure is a point mass, transport has only one possible plan. For kernel distances, the difference between the mean vectors scales by $t$. These observations and the two-point definitions give the formulas in Section~\ref{sec:connections}.

The endpoint term $0\log0$ in~\eqref{eq:contam5} is understood as zero. For example,~\eqref{eq:contam2} follows from
$W_2(\delta_{\theta^\top x},(1-t)\delta_{\theta^\top x}+t\delta_{-\theta^\top x})^2=4t(\theta^\top x)^2$
and $\int(\theta^\top x)^2\dd\sigma=1/d$. In~\eqref{eq:contam3}, the signed difference is $t(\delta_x-\delta_{-x})$ and the endpoint normalized norm distances are one.

Expanding $1-\sqrt{1-t}$ at zero gives $t/2+O(t^2)$. Expanding~\eqref{eq:contam5} gives $\JS=(\log2)t/2+O(t^2)$. Taking the square roots, with the normalization defining $J$, gives the stated common asymptotic behavior of $H$ and $J$.
\end{proof}

\begin{proof}[Proof of Proposition~\ref{prop:topology}]
Weak metrization was proved for the transport, angular, Gaussian MMD, and energy metrics. Probability measures on a compact metric space are weakly compact. In particular, tightness is automatic, so the compactness theorem for probability measures gives sequential compactness. Every compact metric space is complete and separable. Separability can also be seen directly. Finitely supported measures on a countable dense subset, with rational weights, form a countable weakly dense collection.

For TV completeness, the finite probability measure $r=\sum_{n\geq1}2^{-n}\mu_n$ dominates any TV-Cauchy sequence $\mu_n$. Its densities form an $L^1(r)$-Cauchy sequence, whose limit is nonnegative and has integral one. This density defines the TV limit. The uniform two-sided comparisons~\eqref{eq:Htv} and~\eqref{eq:Jtv} transfer both the Cauchy and convergence properties to Hellinger and $J$, proving completeness. For each of these three metrics, the uncountably many point masses have pairwise distance one, ruling out separability. Their convergence implies weak convergence by~\eqref{eq:tvtransport}, while the converging-point-mass example proves strictness.
\end{proof}

\subsection{Finite computations and approximations}\label{app:approx}

\begin{proof}[Equal-weight finite transport]
Multiplying the transport matrix by the common sample size gives a matrix whose row and column sums are one. Every such matrix is a convex combination of permutation matrices, and a linear objective reaches its minimum at one of those permutations.
\end{proof}

\begin{proof}[Proof of Proposition~\ref{prop:pullback}]
Nonnegativity, symmetry, and the triangle inequality follow from the metric on the summaries. Zero distance means that the summaries agree, which forces the original objects to agree exactly when $T$ is one-to-one.
\end{proof}

\begin{proof}[Failure of separation for finite mean features]
A fixed finite vector of average features cannot distinguish \emph{all} probability measures on the sphere. To see this, let $\phi(x)\in\R^m$ and choose $m+2$ distinct points. The vectors $(1,\phi(x))$ lie in $\R^{m+1}$ and are therefore linearly dependent. The first coordinate forces the coefficients of any nonzero dependence to sum to zero. Separate the positive and negative coefficients, then divide each group by their common total. The resulting probability measures differ but have the same average feature vector. The limitation therefore applies to every fixed finite choice of features.
\end{proof}

\begin{proof}[Finite sliced distance and isotropic normalization]
The square root is a norm distance between the selected quantile functions, so it always gives a pseudometric.

The finite quantile vector in Section~\ref{sec:approx} satisfies
\[
 \|\Phi_\mu^L\|^2
 =d\sum_{\ell=1}^Lw_\ell\int(\theta_\ell^\top x)^2\dd\mu(x)
 =d\int x^\top\left(\sum_{\ell=1}^Lw_\ell\theta_\ell\theta_\ell^\top\right)x\dd\mu(x)=1
\]
under~\eqref{eq:tightframe}. The norm and angle metrics on these unit vectors give the two stated pseudometrics by Proposition~\ref{prop:pullback}. For an equally weighted orthonormal basis, the matrix sum in~\eqref{eq:tightframe} is $I_d/d$. Weighted mixtures preserve that identity.
\end{proof}

\begin{proof}[Isotropy without separation]
Both coordinate distributions agree, although the measures differ. Equal weights on the two coordinate directions satisfy~\eqref{eq:tightframe}. The same example extends to higher dimensions by adding zero coordinates and using all coordinate directions with equal weights.
\end{proof}

\begin{proof}[Failure of angular normalization for arbitrary directions]
Arbitrary directions can make the angular formula undefined over the reals. With one direction $\theta$ and measures $\delta_\theta,\delta_{-\theta}$, we have $\SWL^2=4$. Then $1-d\SWL^2/2=1-2d<-1$, which lies outside the domain of $\arccos$. Clipping the argument to $[-1,1]$ does not establish a triangle inequality. Alternatively, one can normalize each nonzero representation to unit length and compare angles. This gives a pseudometric wherever the normalization is defined, but separation still needs to be checked.
\end{proof}

\begin{proof}[Convergence of directional averages]
For fixed $\mu,\nu$, the projection coupling bound makes $\theta\mapsto W_2^2(\mu_\theta,\nu_\theta)$ bounded and continuous. Weak convergence of $\lambda_L$ therefore gives $\SWL^2(\mu,\nu)\to\SW^2(\mu,\nu)$.
\end{proof}

\begin{proof}[Unbiased MMD estimator and its negative example]
Each off-diagonal within-sample pair has the expectation of an independent pair from the same population. Each cross-sample pair has the corresponding cross-population expectation. Linearity of expectation therefore gives population MMD squared. In the two-point example, the two within-sample terms are both $a$, while the cross term is $1+a$. Their difference is $a-1<0$, as stated in Section~\ref{sec:approx}.
\end{proof}

\subsection{Squared distances and Sinkhorn divergences}\label{app:sinkhorn}

\begin{proof}[The three-point squared-distance counterexample]
Take point masses at angles $0,\pi/3,2\pi/3$ on a great circle. Squared chordal distances are $1,1,3$ for the two adjacent pairs and the outer pair, so the triangle inequality would require $3\leq2$. Squared geodesic distances give $(2\pi/3)^2>2(\pi/3)^2$. On point masses, the sliced squared distance is squared chordal distance divided by $d$ and fails for the same reason. Since $1-\Cot=W_{2,\mathrm c}^2/2$ and $1-\Csw=d\SW^2/2$, both unrooted cosine dissimilarities also fail. The mixture counterexamples in Sections~\ref{sec:energy} to~\ref{sec:js} treat squared MMD, energy, squared Hellinger, and JS.
\end{proof}

\begin{proof}[Regularized transport and Sinkhorn divergence]
Independent pairing has zero penalty, and $c\leq1$, giving $0\leq T_\varepsilon\leq1$. In fact, Theorem~1 of \citet{feydy_2019_InterpolatingOptimalTransport} applies because the cost is Lipschitz on the compact sphere and $e^{-c/\varepsilon}$ is a positive universal Gaussian kernel. A Lipschitz function has changes in value bounded by a constant times changes in its inputs. Universality means that the associated Hilbert space of functions can approximate every continuous function uniformly.

Universality follows from the signed-measure version of~\eqref{eq:gaussfourier}. Otherwise, the Hahn-Banach and Riesz representation theorems would give a nonzero finite signed measure integrating every kernel function $x\mapsto e^{-c(x,y)/\varepsilon}$ to zero. Its Gaussian kernel embedding would then vanish, contradicting the earlier argument. The cited theorem gives $S_\varepsilon\geq0$, with equality exactly for equal measures. Because the subtracted self-costs are nonnegative, we obtain
the stated upper bound on $S_\varepsilon$.
The self-cost is positive unless the measure is a point mass. A minimizing coupling exists because the coupling set is compact, the objective is lower semicontinuous, and independent pairing has finite objective. A zero minimum would require both zero transport cost and zero KL penalty. Zero penalty forces independent pairing, while zero cost forces the paired points to agree almost surely. Both conditions hold only for a point mass.
\end{proof}

\begin{proof}[Verification of Example~\ref{ex:sinkhorn}]
The symmetric two-by-two transport problem has diagonal cost $a=\sin^2((s-t)/2)$ and off-diagonal cost $b=\sin^2((s+t)/2)$. Writing $q$ for the total diagonal mass gives the objective
\[
 qa+(1-q)b+\varepsilon\{q\log(2q)+(1-q)\log(2(1-q))\}.
\]
Minimizing over $q\in[0,1]$ gives the exact formula
\begin{equation*}
 T_\varepsilon(\mu_s,\mu_t)
 =\varepsilon\log2-\varepsilon\log(e^{-a/\varepsilon}+e^{-b/\varepsilon}).
\end{equation*}
When $s=0$, the first measure is a single point and both costs agree. The formula then equals their common cost.

Put $L(z)=\log(1+e^{-z})$ and abbreviate $S_{ij}=S_\varepsilon(\mu_{i\theta},\mu_{j\theta})$. Exact trigonometric identities give
\begin{align*}
 S_{01}&=0.02-0.01\log2+0.01L(3.92),\\
 S_{02}&=0.0784-0.01\log2+0.01L(14.450688),\\
 S_{12}&=0.02+0.01L(3.92)+0.01L(14.450688)-0.02L(7.5264).
\end{align*}
All terminating decimals here denote exact rationals. The elementary bounds
$0.69<\log2<0.70$, $e^{-3.92}<0.025$, $e^{-14.450688}<0.001$, and $0<L(z)<e^{-z}$ imply
\[
 S_{01}<0.01335<0.116^2,\qquad
 S_{12}<0.02026<0.143^2,\qquad
 S_{02}>0.0714>0.267^2.
\]
Thus
$\sqrt{S_{02}}>0.267>0.116+0.143>\sqrt{S_{01}}+\sqrt{S_{12}}$.
These strict bounds prove the triangle-inequality failure without numerical rounding. Adding zero coordinates gives the same example in every $d\geq2$. Figure~\ref{fig:sinkhorn-counterexample} shows the distributions and compares the direct score with the sum through the intermediate distribution.
\end{proof}

\end{document}